\documentclass[12pt]{elsarticle}
\pdfoutput=1
\usepackage{lineno,hyperref,color}
\modulolinenumbers[5]

\journal{Journal of \LaTeX\ Templates}

\begin{document}

\begin{frontmatter}

\title{Discrimination of anisotropy in dark matter velocity distribution with directional detectors}

\author{Keiko I. Nagao}
\address{Faculty of Fundamental Science, National Institute of Technology, Niihama College,
\\Niihama, Ehime 792-8580, Japan,

\& Faculty of Science, 
Okayama University of Science, \\
Okayama, Okayama 700-0005, Japan}
\author{Tomonori Ikeda}
\address{Department of Physics, Kobe University, Kobe, Hyogo 657-8501, Japan}
\author{Ryota Yakabe}
\address{Department of Physics, Kobe University, Kobe, Hyogo 657-8501, Japan}
\author{Tatsuhiro Naka}
\address{Department of Physics, Faculty of Science, Toho University,
Funabashi,Chiba 274-8501, Japan,

\& Kobayashi-Maskawa Institute, Nagoya University,
Nagoya, Aichi 464-8601, Japan}
\author{Kentaro Miuchi}
\address{Department of Physics, Kobe University, Kobe, Hyogo 657-8501, Japan}

\begin{abstract}
The directional detection of dark matter is sensitive to both recoil energy and the direction of nuclear recoil.
It provides a technique whereby the local velocity distribution of dark matter may be measured. 
In this study, the possibility of discriminating between isotropic and anisotropic distributions is investigated 
through numerical simulations with a directional detector.
The numerical simulation
is performed for two cases. These cases are classified according to detectors as follows: 
one corresponds to an angular histogram distribution of the signals whereas the other corresponds to an
energy-angular distribution of the signals. In order to discriminate the energy-angular distributions,
a chi-squared test between ideal data set and experimental data and likelihood estimation are proposed.
The anisotropy of the velocity distribution was shown to be discriminated at 
$90$\% confidence level if O($10^4$) signals are obtained 
for both gaseous detector and solid detector. Especially, the analysis for the solid detector case and comparison with the gaseous detector case are discussed for the first time.
\end{abstract}


\end{frontmatter}


\section{Introduction}
Recent astronomical and cosmological observations have revealed that non-luminous 
matter accounts for about 85\% of the matter in the Universe. This is known as the dark matter \cite{Planck}. 
Weakly Interacting Massive Particles (WIMPs) have been suggested as promising candidates for dark matter. 
Direct searches of WIMPs have been performed, and a review of these results is documented in the literature \cite{Drees}. 
In direct detection experiments, the target is expected to be scattered by 
WIMPs and leave the recoil energy. Since environmental and astronomical background signals are 
also detected in these experiments, these contributions must be removed to quantify the signal due to 
dark matter. This rejection of background signal contributions is a  highest priority issue because of  
the small cross section of WIMPs. Directional detection experiments are sensitive to both recoil energy 
and direction \cite{Ahlen, Mayet2016}.
On account of the revolution of the Solar system within the Galaxy, dark matter signals are expected to come from the 
direction in which the Solar system is traveling, namely, from the direction of Cygnus. By comparison, direction of 
environmental background signals is assumed to be isotropic over the sky. The use of the difference in  directionality
of these signal sources will facilitate the rejection of spurious background signals.

The study is now in progress \cite{Ahlen,Battat} on gaseous time projection chamber (TPC) and solid detector (e.g., the fine-grained nuclear emulsions \cite{Aleksandrov, Naka}, carbon-nanotube and anisotropy crystal scintillator).
The gaseous TPC technique detects 3D tracks due to WIMPs with track length in the sub-millimeter regime. 
The resolution is dependent on both the readout system of the TPC and diffusion.
For solid state detectors, the typical track length of WIMPs is less than a submicron. 
As a result, very high-resolution detectors are required to distinguish signals as tracks. 
Fine-grained nuclear emulsions have demonstrated the ability to detect tracks of very short lengths.
This technique is able to record 3D tracks greater than 50 nm, but the performance depends on the associated microscopy system employed.
As stated above, the study of directional dark matter detectors is a very active field of research  
and an investigation into the application of these unique techniques will have a substantial impact on the development of future dark matter astronomy.

The velocity distribution of dark matter is the primary subject of this work,
and in most cases, an isotropic Maxwell-Boltzmann distribution is adopted to describe the velocity distribution \cite{LewinSmith}.
Some studies have included non-Maxwellian distributions; therefore, these studies deviate from the above simple model. 

In current hydrodynamical simulations, the velocity distribution should not necessarily be Maxwell-Boltzmann distribution \cite{Bozorgnia:2017brl}.
For example, some
numerical simulations indicate non-Maxwellian distributions
due to a dark disk [9-11], tidal streams \cite{stream}, and debris flows \cite{debris,KLS},
and some of them indicate anisotropic 
velocity distributions. 
Deviation from an isotropic Maxwellian distribution slightly shifts 
the result of ordinary direct detections \cite{Shan}--\cite{Mao:2013nda}. In the directional detection experiments, 
the difference can be expected to be more obvious than that in the ordinary detections since both recoil energy 
and the directional information can be used \cite{OHare, Kavanagh, Kavanagh:2016xfi}. 
The discrimination between the isotropic and anisotropic distribution of dark matter in the directional detection experiments is
discussed in literatures \cite{Morgan:2004ys,Host:2007fq} for the case of the gasses detector. In this work, we revisit it with the chi-squared test approach not only for the gasses detector but also for the solid detector.

This study is organized as follows: In Section 2, a short review of the velocity distribution and setup
of the numerical simulation is given. Analysis of the calculation performed is given in Section \ref{sec:detailsoftheanalysis}
and Section \ref{sec:rdetermination}. 
Two cases are discussed as defined by the angular/energy resolution of the detector: One is an energy-angular distribution and the
other is the angular histogram.
Conclusions of these simulations are presented in Section \ref{sec:conclusion}.
\section{Direct Detection with Directional Detectors}
\label{sec:setting}
In most cases, an isotropic velocity distribution is commonly supposed to derive constraints from
direct detections. The Maxwell-Boltzmann distribution \cite{LewinSmith}
is a typical distribution function; however, N-body simulations \cite{LNAT, KLS} and observations \cite{Sagittarius} 
suggest the presence of an anisotropic component in the distribution that can influence the result of direct detection experiments.
The effect of this factor on conventional direct detections \cite{Shan}--\cite{Hunter:2013vua} and directional 
detections \cite{Billard:2009mf}--\cite{Host:2007fq} has been the subject of a number of studies. 
In this work, the anisotropic distribution function of N-body simulations  \cite{LNAT} is used in our numerical simulations. 
Note that the distribution is adopted as a benchmark scenario of anisotropic distribution, 
and not necessarily realized.
The tangential velocity of dark 
matter with respect to the galactic rest frame $v_\phi$ has the distribution:
\begin{eqnarray}
f(v_\phi) = \frac{1-r}{N(v_{0,\mathrm{iso.}})} \exp\left[-v_\phi^2/v_{0,\mathrm{iso.}}^2\right] + 
		\frac{r}{N(v_{0,\mathrm{ani.}})} \exp\left[-(v_\phi-\mu)^2/v_{0,\mathrm{ani.}}^2\right] ,
\label{eq:doublegaussian}
\end{eqnarray}
where the normalization factor $N(v_0) = 2v_0 \Gamma(3/2)$, $v_{0,\mathrm{iso.}} = 250$ km/s, 
$v_{0,\mathrm{ani.}} = 120$ km/s and $\mu=150$ km/s. The fraction factor of the double Gaussian component $r$ is $0.25$
in \cite{LNAT}, and calculations are performed with the parameter as $r = 0.0, 0.3, 1.0$ account for 
changes in the degree of anisotropy. 
It is easily understood that the parameter $r$ corresponds to 
percentage of anisotropic component in the total velocity distribution.
In this simulation, the distributions of radial velocity $v_r$ and velocity across the galactic plane $v_z$ 
are also defined as Gaussian and are described as follows:
\begin{eqnarray}
f(v_r) = \frac{1}{N(v_{0,r})} \exp\left[-v_r^2/v_{0,r}^2\right] \\
f(v_z) = \frac{1}{N(v_{0,z})} \exp\left[-v_z^2/v_{0,z}^2\right] 
\label{eq:velocitydistribution}
\end{eqnarray}
where $v_{0,r}=240.4$ km/s and $v_{0,z}=214.6$ km/s.

\begin{figure}[t!] 
   \centering
   \includegraphics[width=2.5in]{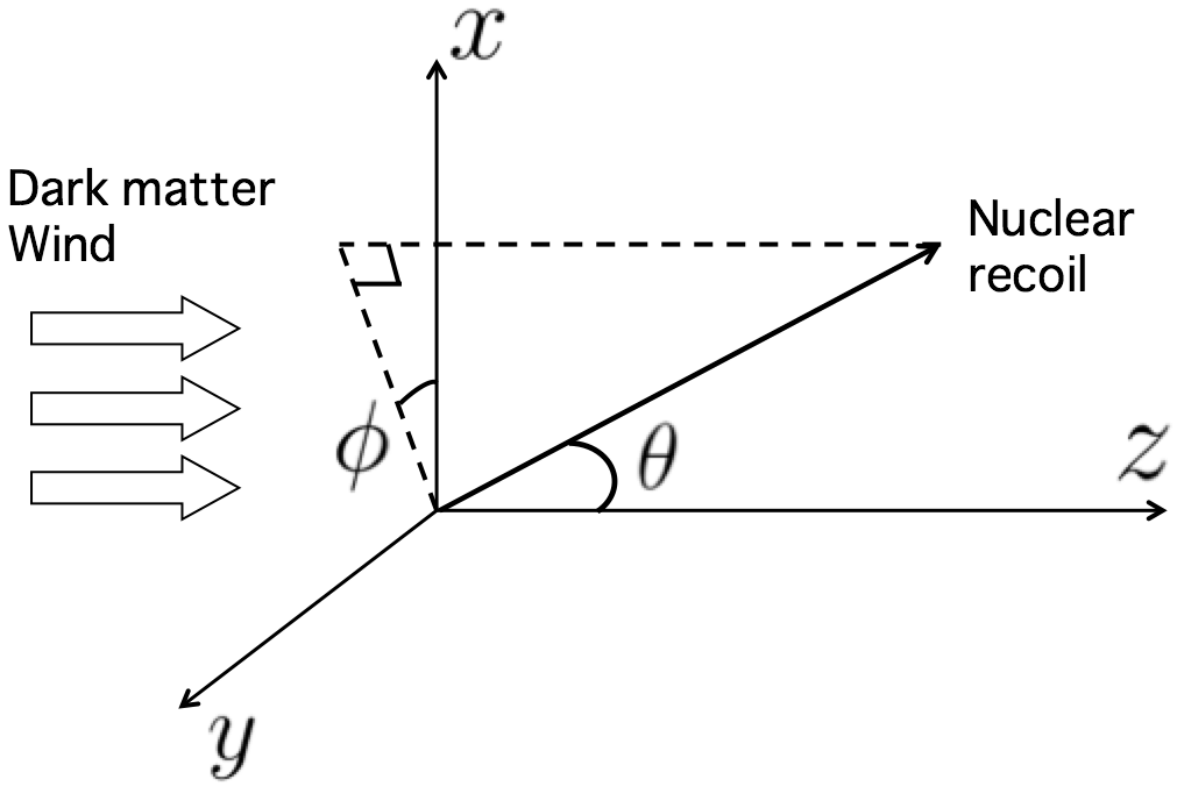} 
   \caption{Nuclear recoil in the laboratory frame. The scattering angles of the nuclear recoil are taken as $(\theta, \phi)$. 
   }
   \label{fig:labsys.eps}
\end{figure}

In this calculation, a  Monte-Carlo simulation of the dark matter scattering was performed. 
In the Galactic rest frame, velocity distribution was assumed. WIMPs are generated with the velocity following the distribution. 
After converting the velocity to that in the laboratory rest frame, scattering of nuclei and WIMPs are simulated.
The scattering angles with respect to the laboratory frame are shown in Figure. \ref{fig:labsys.eps}. 
The scattering angle
of the nucleon from the direction of the dark matter wind toward the Earth and its tangential angle are $\theta$ and $\phi$, 
respectively. 
By the simulation, data set of recoil energy $E_R$, scattering angle  $\theta$ and $\phi$ are obtained.In the following analysis, 
a detector which has enough sensitivity to measure 3-dimensional direction is supposed and only $\theta$ is used as the scattering angle 
because $\phi$ is not affected by the velocity distribution.

\begin{table}[t]
\begin{center}
  \begin{tabular}{|c|c|c|} \hline
    target &  mass number & interaction \\ \hline \hline
    F & 19.00 & SD \\ \hline
    Ag & 107.87 & SD \\ \hline
    C & 12.01 & SI \\ \hline
    S & 32.06 & SI \\ \hline
    Br & 79.90 & SI \\ \hline
    I & 126.90 & SI \\ \hline
  \end{tabular}
  \label{tab:target}
  \caption{Typical target atoms and associated parameters used in directional experiments. 
  Spin-dependent interactions and spin-independent interactions are represented by SD and SI, respectively. }
  \end{center}
\end{table}

Common target in directional detector are carbon (C), F (fluorine), S (sulfur) for the gaseous detector 
while C, bromine (Br), silver (Ag) and iodine (I) are employed in solid state detectors. 
Parameters associated with typical target atoms employed in directional experiments are found in Table~1. 
In the calculation presented in this work, F was chosen as a typical light target element
and Ag as a typical heavy target element. Thus $m_A=17.7$ GeV and $100.6$ GeV for Fluorine and Silver target case, respectively.
As described previously, form factors for F and Ag are assumed to be given by Eq.(4.2) and Eq.(4.3) in \cite{LewinSmith}, respectively.
Elastic scattering is also assumed in this simulation.
The recoil energy and the scattering 
angle depend only on the masses of dark matter and  that of the target atom. Thus, we can refer target with near mass as target of interest. 
A mass ratio of $m_\chi/m_A=3$ is assumed for simplicity
together with a zero background signal and ideal detector resolution. 

\section{Analysis Method}
\label{sec:detailsoftheanalysis}

Both the recoil energy and the direction of the nuclear recoil are detected with a typical directional detector. 
Therefore, these directional detectors can detect both the recoil energy and the scattering angle.
Discrimination between isotropic and
anisotropic distributions generally can be realized with a two-dimensional energy and angle distribution. 
This standard method is discussed in Section \ref{subsec:ERcostheta} and Section \ref{subsec:chi2_distribution}.
Some directional detectors do not have a very good energy resolution or threshold-type detectors.
In these cases, only the angular distribution
of the nuclear recoil can be used to test the anisotropy. 
This case is discussed in Section \ref{subsec:costhetahist} and Section \ref{subsec:chi2_histogram}.

In both of the two cases, two sets of numerical simulation result  are presented.
The first set has  a large event number of $O(10^8)$. 
This describes an ideal situation, 
but it is difficult to achieve experimentally. 
This will be referred to as the ideal ``{\bf template}''. 
The second case is for a smaller event number than the template, in the range 
$O(10^3)$--$O(10^4)$. This describes a more realistic situation than that of the ideal template and will be referred to
as the {\bf ``pseudo-experiment''}.
The pseudo-experiment data is compared with template data to discriminate the anisotropic halo models in the following discussions.

\subsection{The energy-angular distribution}
\label{subsec:ERcostheta}

\begin{figure}[t]
 \centering
\includegraphics[keepaspectratio, scale=0.19,clip]{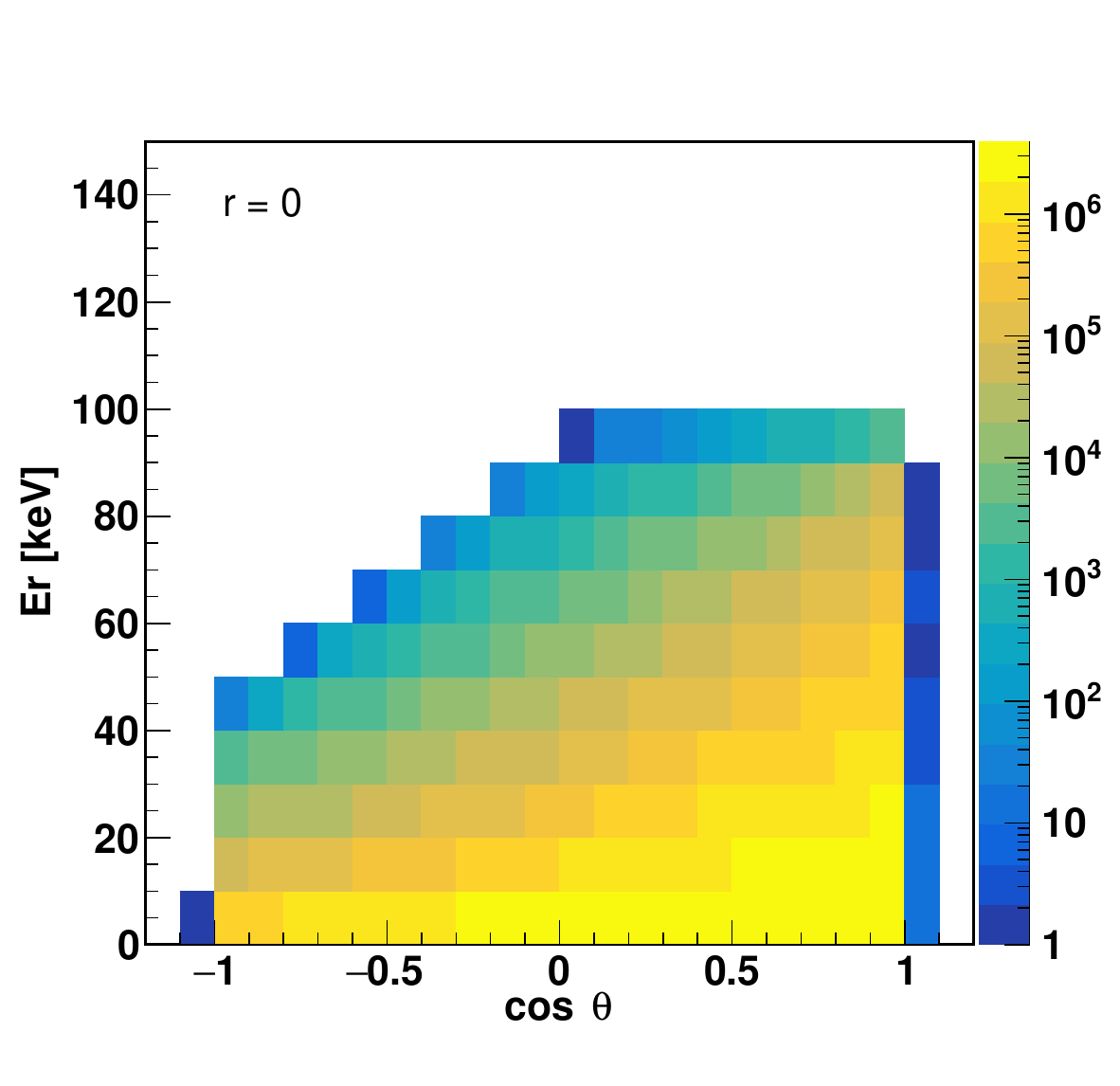}
\includegraphics[keepaspectratio, scale=0.19,clip]{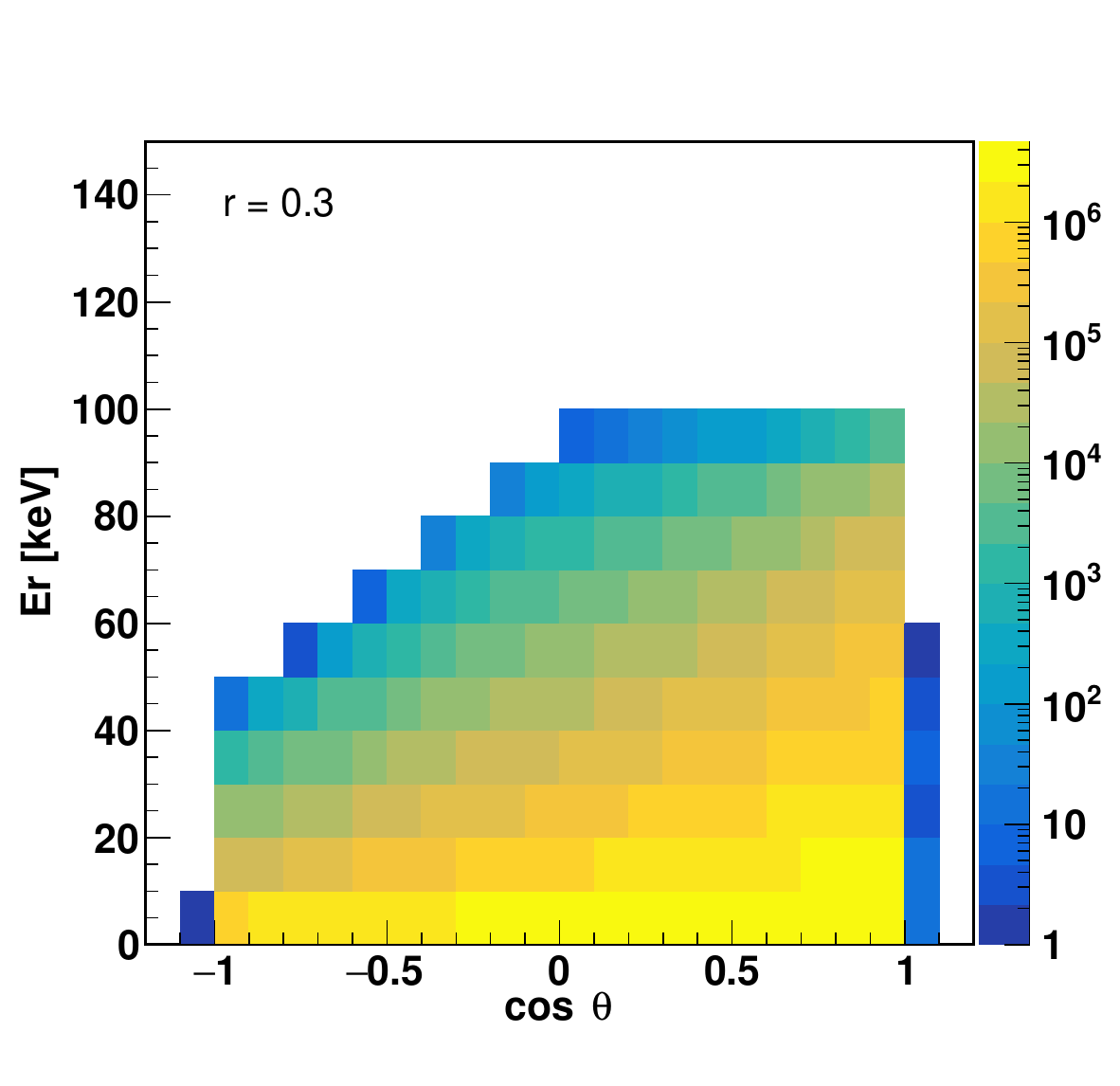}
\includegraphics[keepaspectratio, scale=0.19,clip]{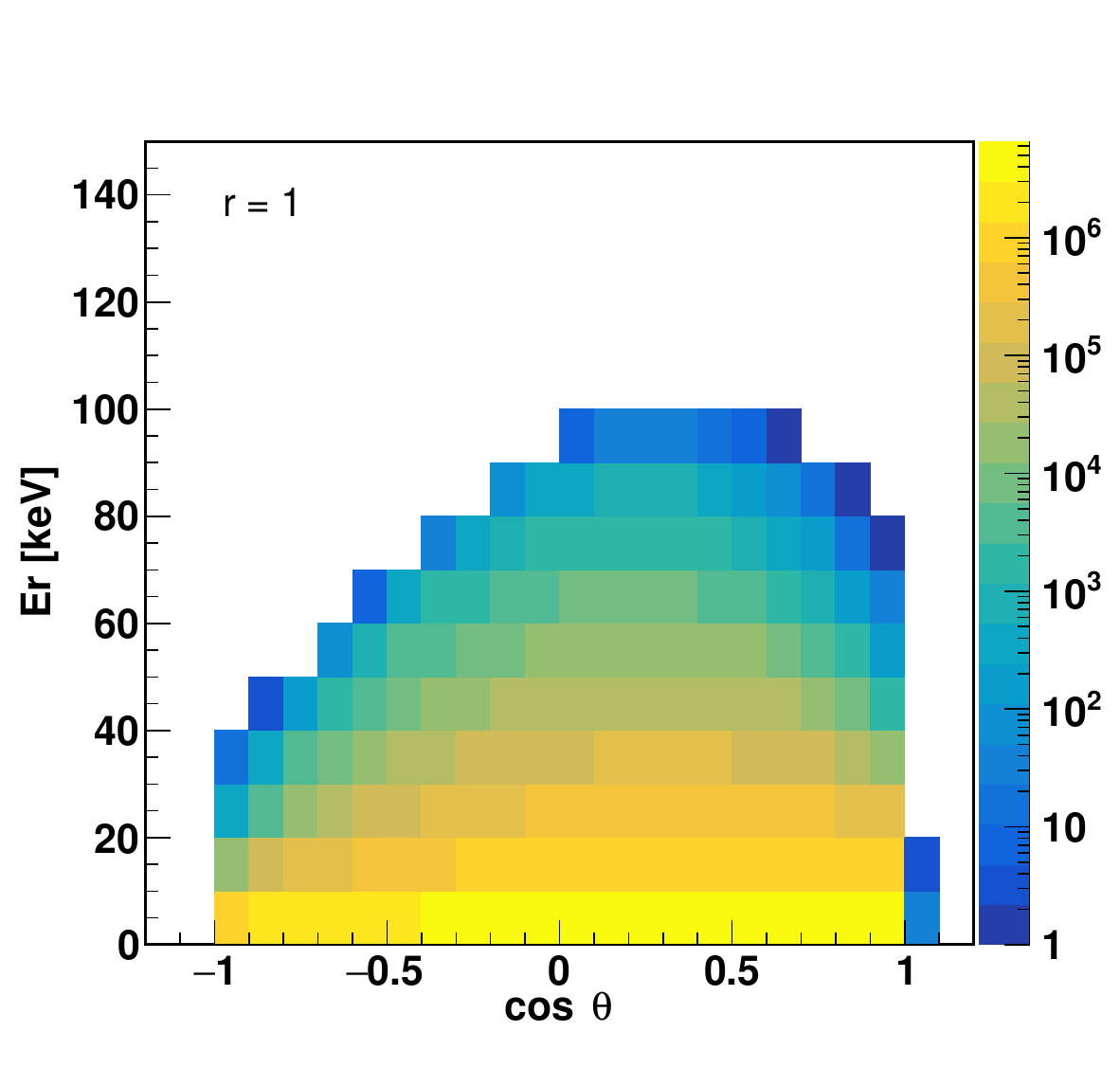}
\caption{$E_R$-$\cos{\theta}$ distribution for the $m\chi=3m_A$ case. The target nucleon is F (light) and 
 the energy threshold $E_R^{\mathrm{thr}}= 0$ keV is assumed in all the figures.}
 \label{fig:ERcos_F}
 \end{figure}
 \begin{figure}[t]
 \centering
\includegraphics[keepaspectratio, scale=0.19,clip]{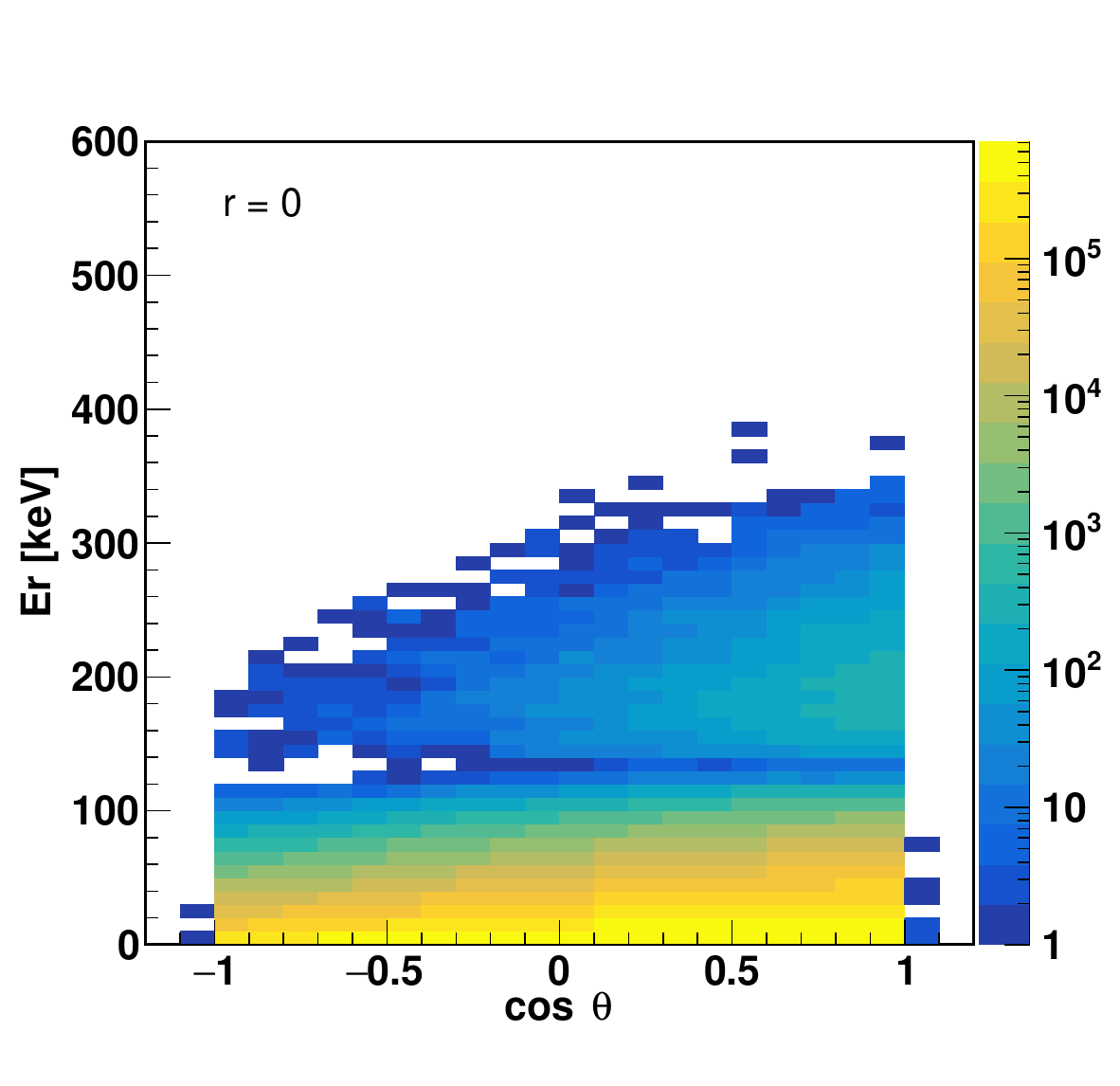}
\includegraphics[keepaspectratio, scale=0.19,clip]{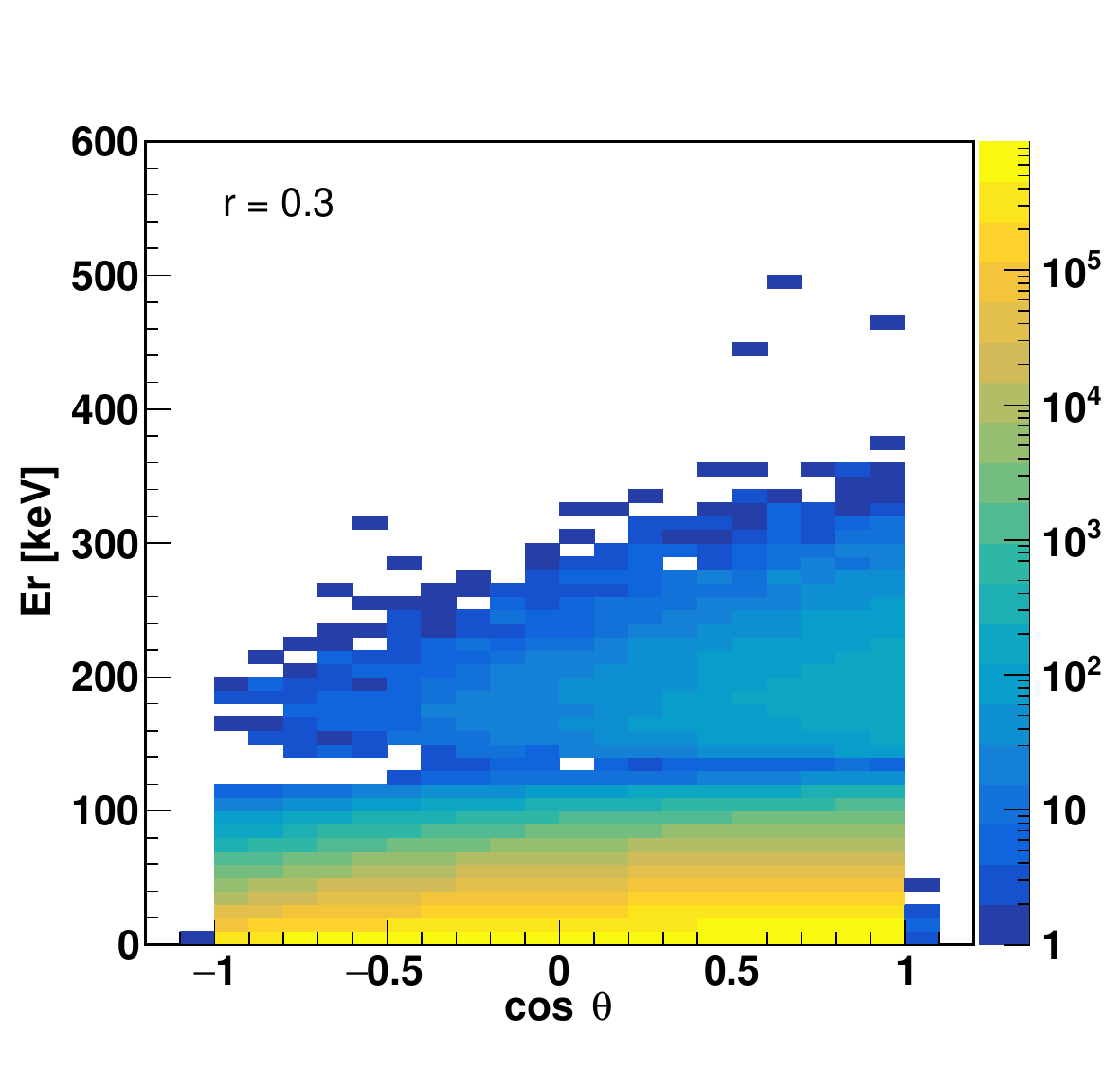}
\includegraphics[keepaspectratio, scale=0.19,clip]{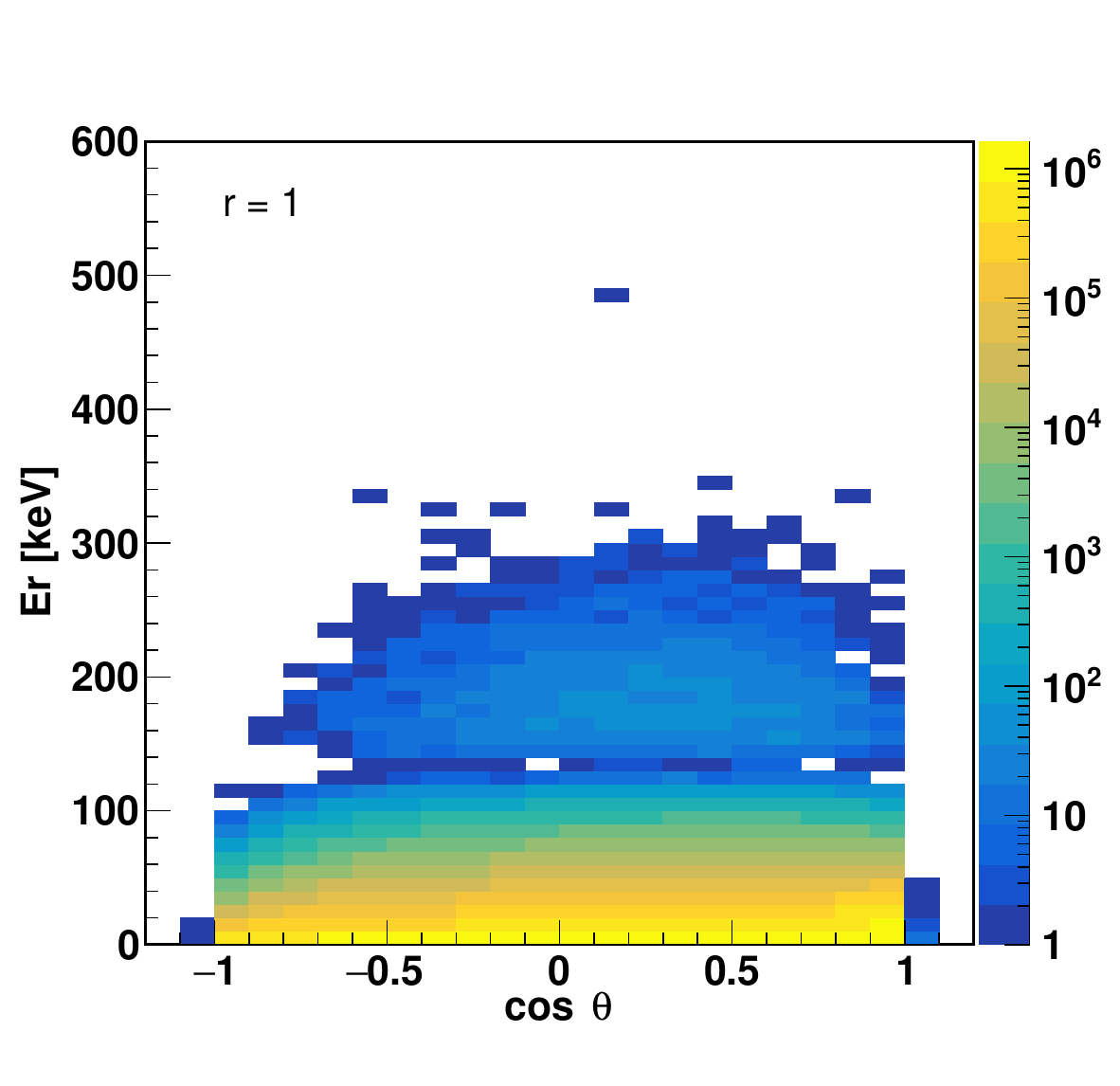}
 \caption{$E_R$-$\cos{\theta}$ distribution for the $m\chi=3m_A$ case. The target nucleon is Ag (heavy) and 
 the energy threshold $E_R^{\mathrm{thr}}= 0$ keV is assumed in all the figures.}
 \label{fig:ERcos_Ag}
\end{figure}

If both the energy resolution and spatial resolution are obtained, analysis of the signal using the recoil energy and
angle is possible. 
In Figures \ref{fig:ERcos_F} and \ref{fig:ERcos_Ag} the $E_R^\mathrm{thr}$-$\cos{\theta}$ distributions of the template data are shown,
where $E_R^\mathrm{thr}$ is the energy threshold of the detector.
Ten values of $r$ in the range of 0 to 1 are assumed in each figure of Figure \ref{fig:ERcos_F} and \ref{fig:ERcos_Ag}. 
Here, $r$ represents the degree of anisotropy defined in Eq.(\ref{eq:doublegaussian}).
The energy-angler distributions down to $0$ keV are shown in the Figures. 
Distributions with realistic detector thresholds can be obtained as subsets of these distributions.
In the figures, most of events condense to low energy region. 
For small $r$ cases, the peak at $\cos\theta \sim1$ shows up as the recoil
energy becomes high since forward scattering events have higher recoil energy.

Isotropic and anisotropic components give contributions to different parameter region of the figures. 
Events caused by the isotropic component  tend to condense in $\cos{\theta}\sim 1$ and has a peak in the region,
while events by the anisotropic component has a peak around $\cos{\theta} \sim 0.4$. It is because the anisotropic component
of dark matter follows the motion of baryons and gasses in the Galaxy, and its velocity with respect to the Earth is lower than
that of isotropic component.  As a result, total distribution of isotropic case ($r\sim 0$) makes the  peak at $\cos\theta \sim 1$, 
on the other hand, that of the anisotropic case ($r=1$) makes the peak around $r\sim 0.4$.
The shape of the distribution depends only on the mass of the dark matter and of the target. However, if the target is 
heavy, the event number in the particular energy region is reduced due to the form factor. 
The shape of the histogram for the Ag target is, therefore, not similar to that of the F target.

\subsection{Angular histogram}
\label{subsec:costhetahist}

\begin{figure}[h]
 \centering
\includegraphics[keepaspectratio, scale=0.19,clip]{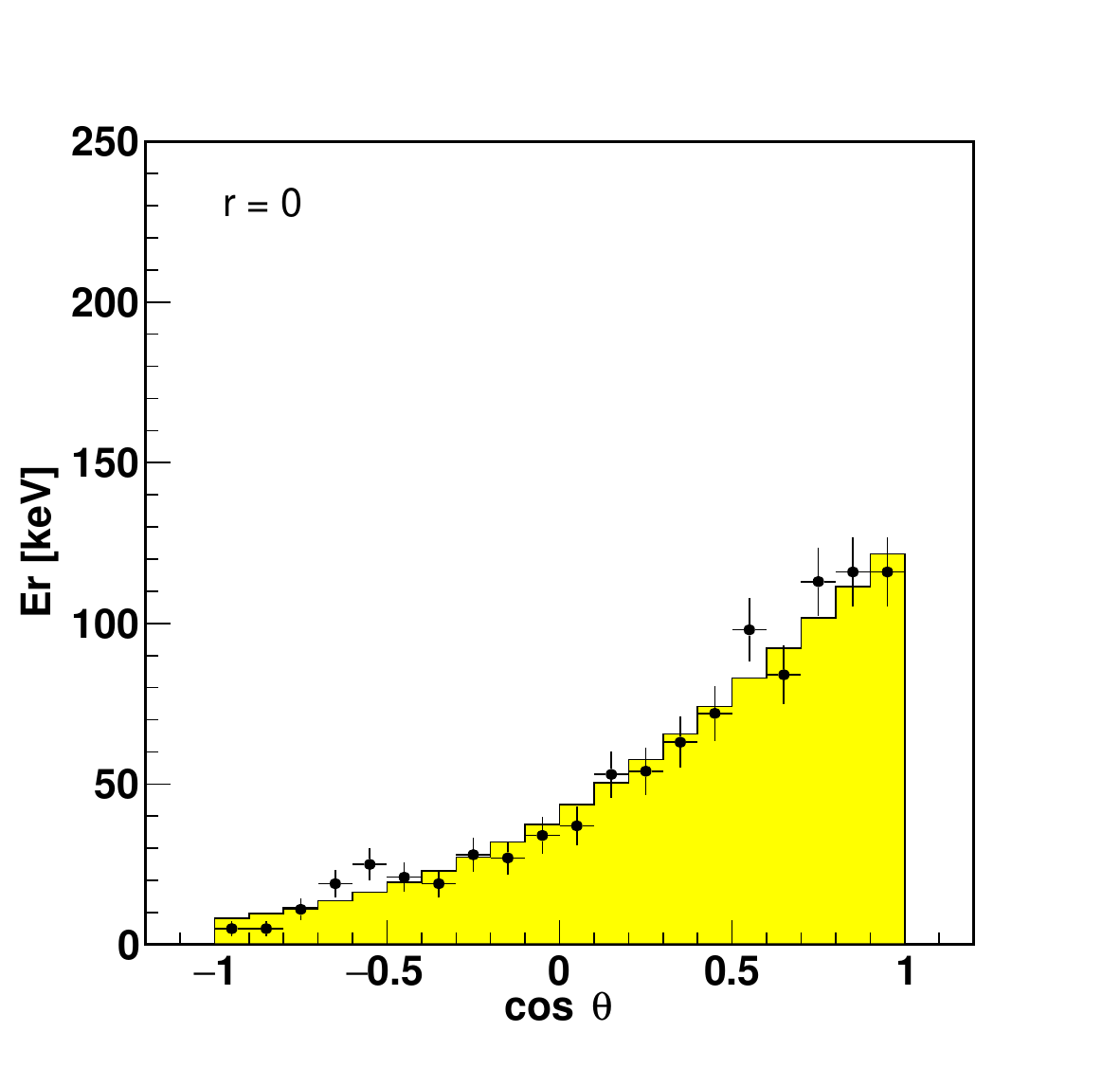}
\includegraphics[keepaspectratio, scale=0.19,clip]{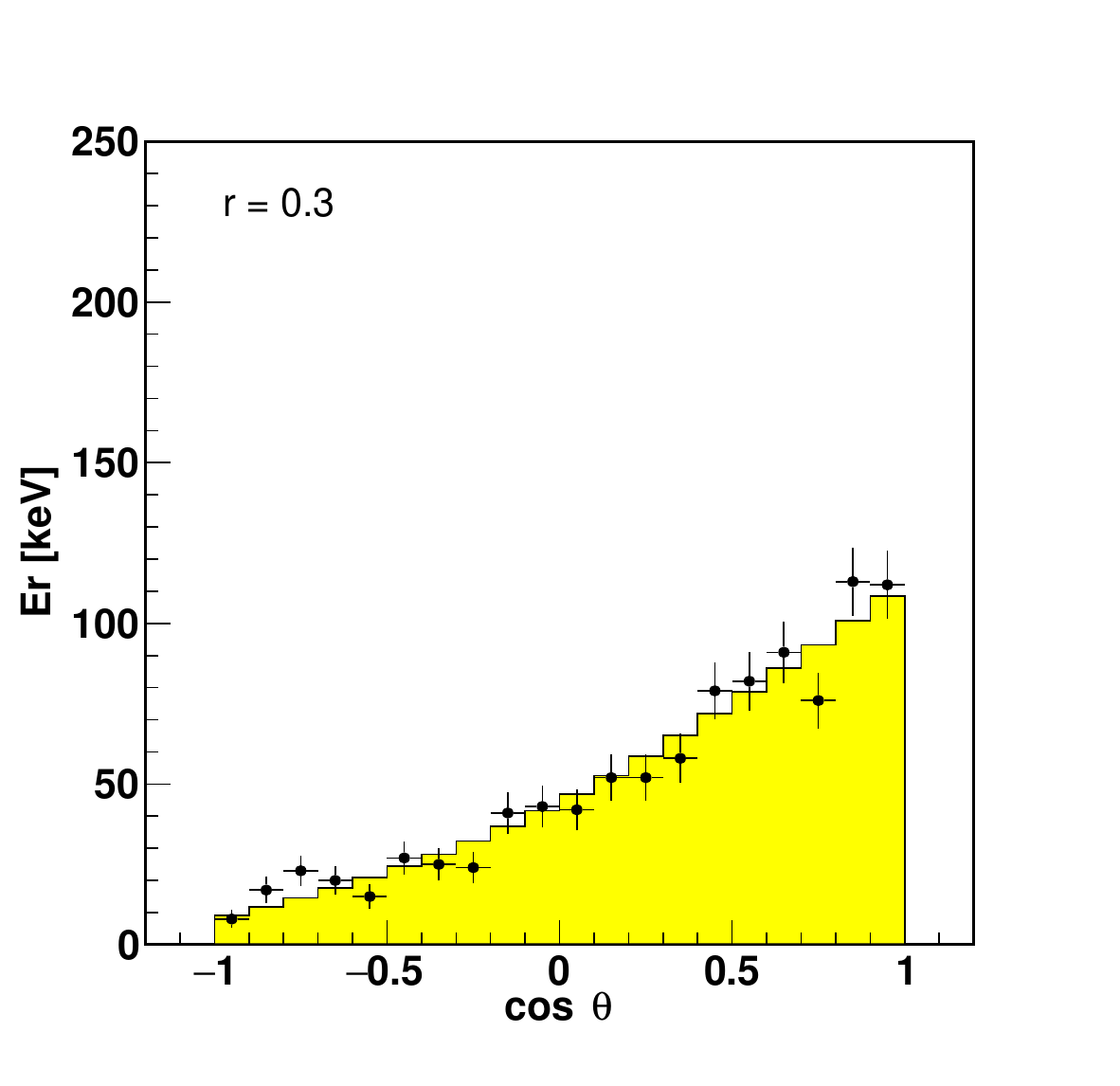}
\includegraphics[keepaspectratio, scale=0.19,clip]{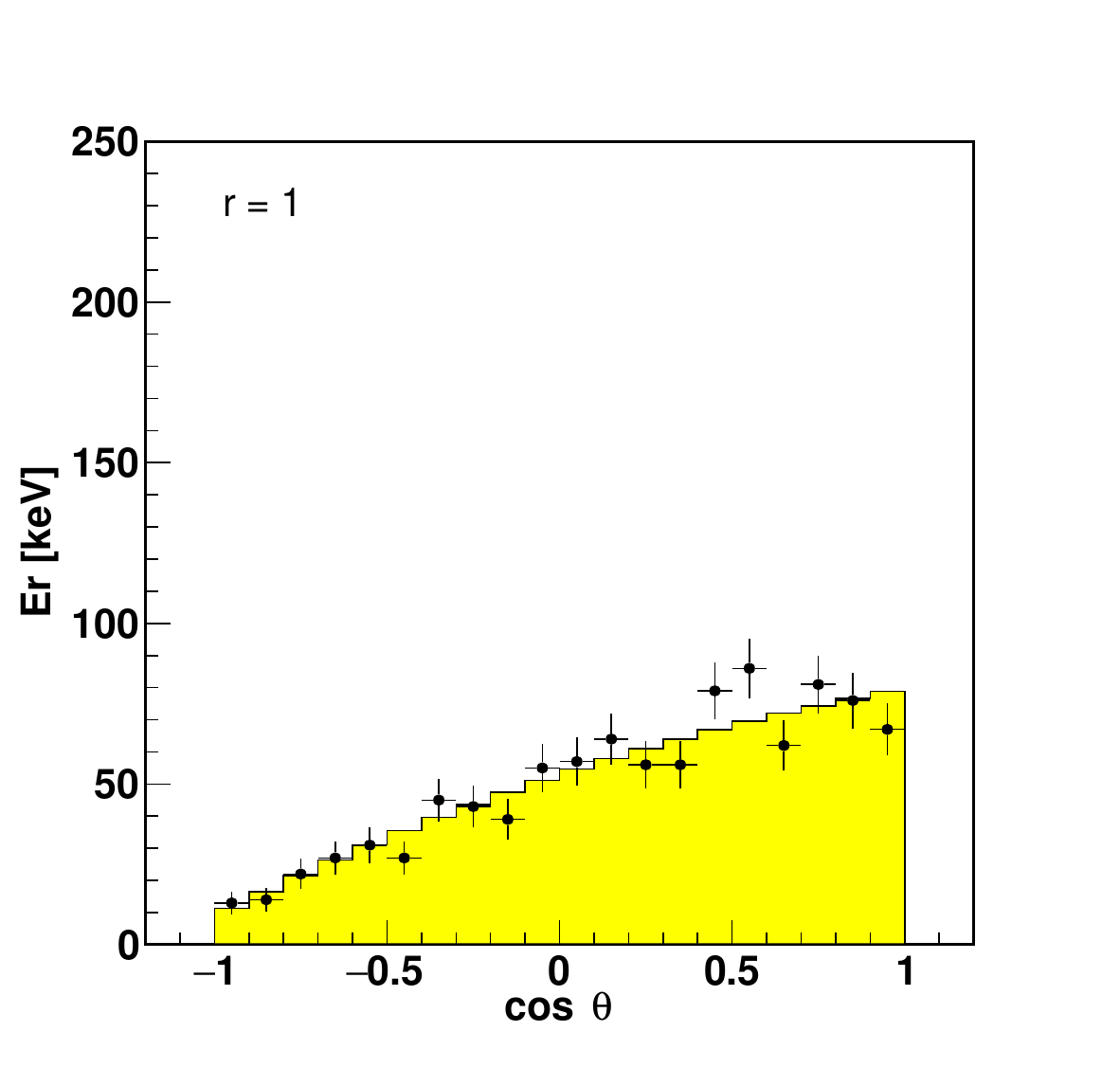}
 \caption{Histogram of $\cos\theta$ the for case of F as the target nucleon (light target), $m\chi=3m_A$, and the recoil energy cutoff  $E_R^{\mathrm{thr}} = 0$ keV. Yellow histograms correspond to the template (event number: $10^8$), and black points represent
 histograms for the pseudo-experiment (event number: $10^3$) together with statistically approximate error bars.}
  \label{fig:coshistF_ER0}
\end{figure}
\begin{figure}[h]
 \centering
\includegraphics[keepaspectratio, scale=0.19,clip]{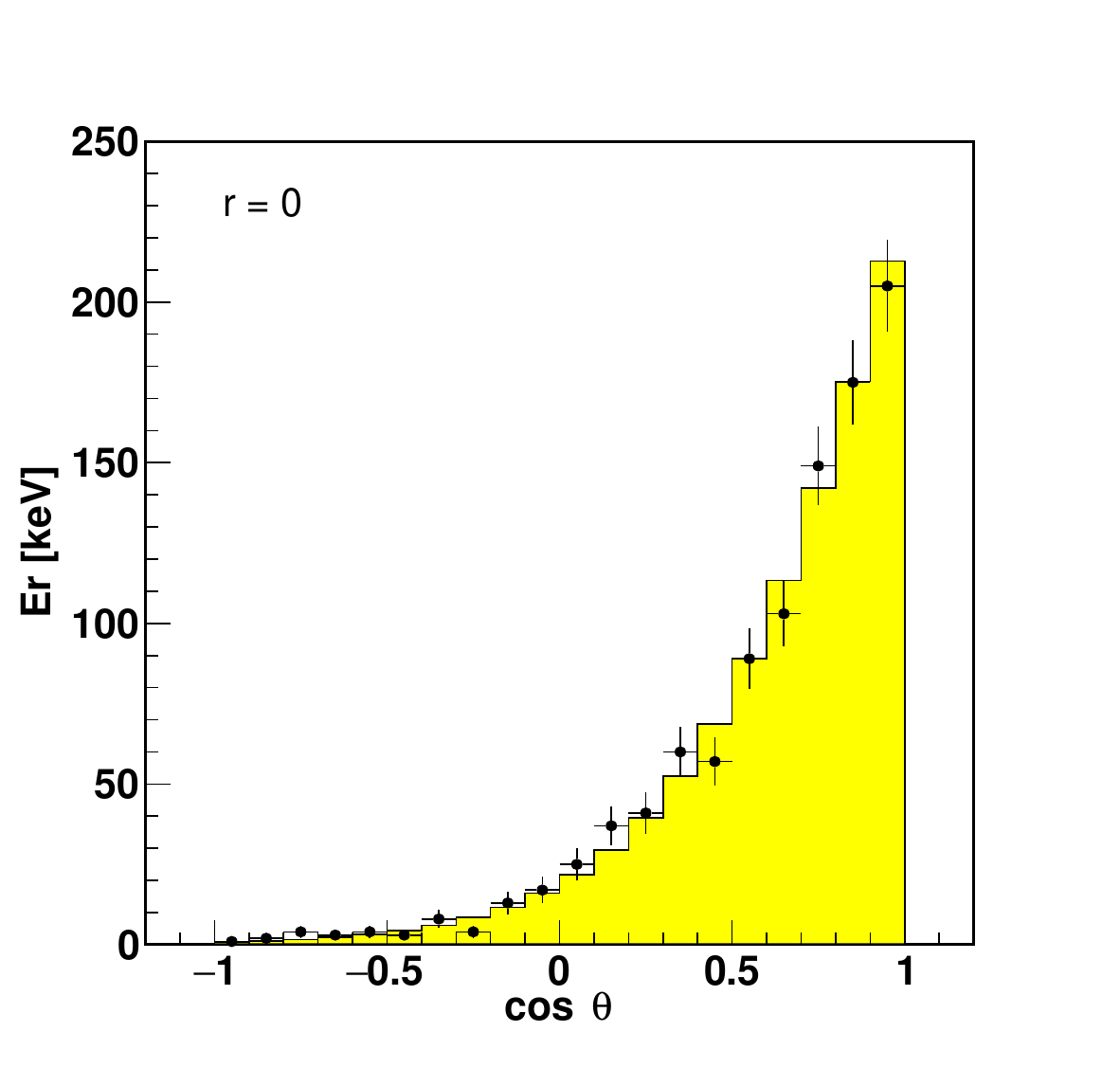}
\includegraphics[keepaspectratio, scale=0.19,clip]{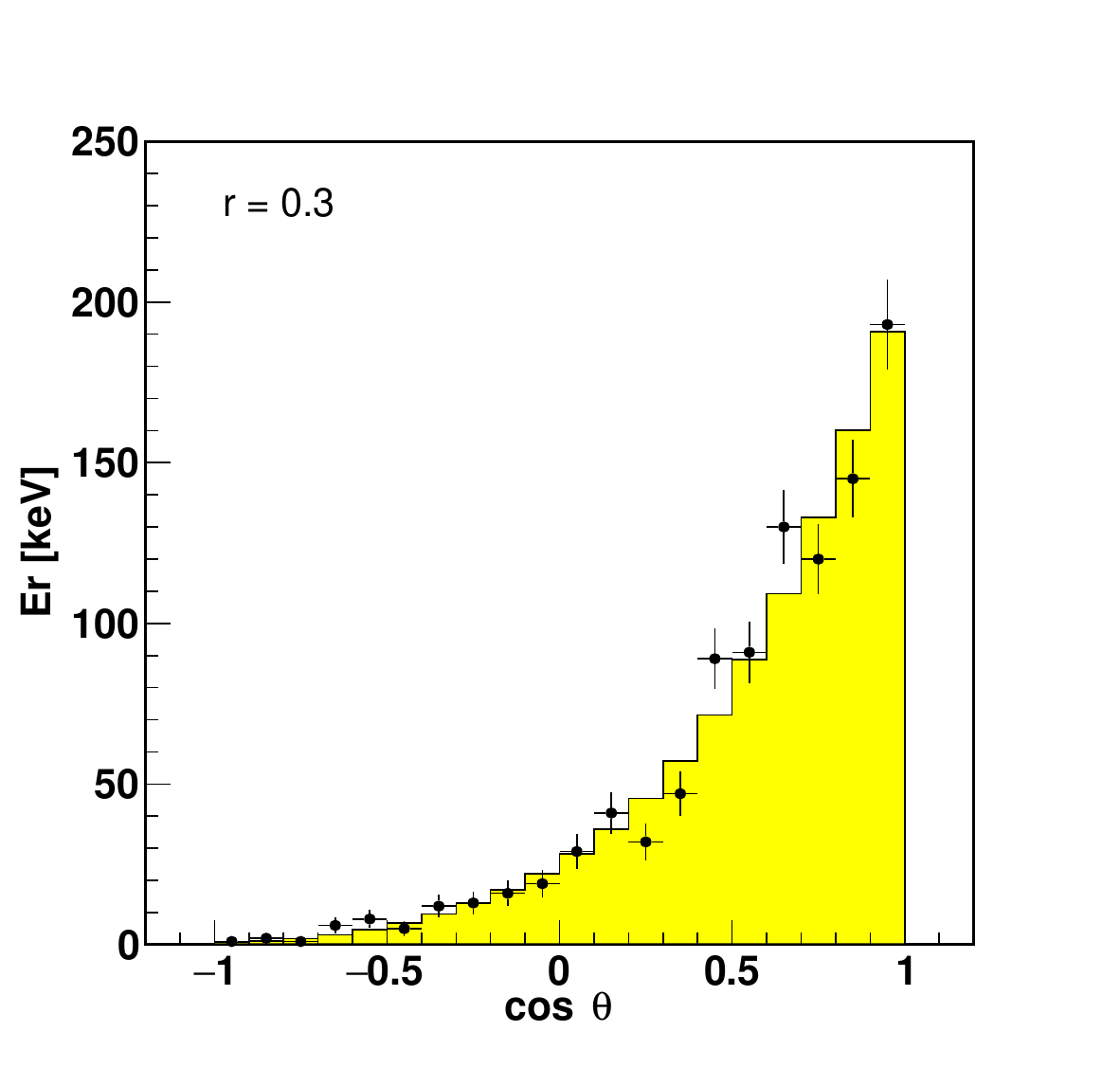}
\includegraphics[keepaspectratio, scale=0.19,clip]{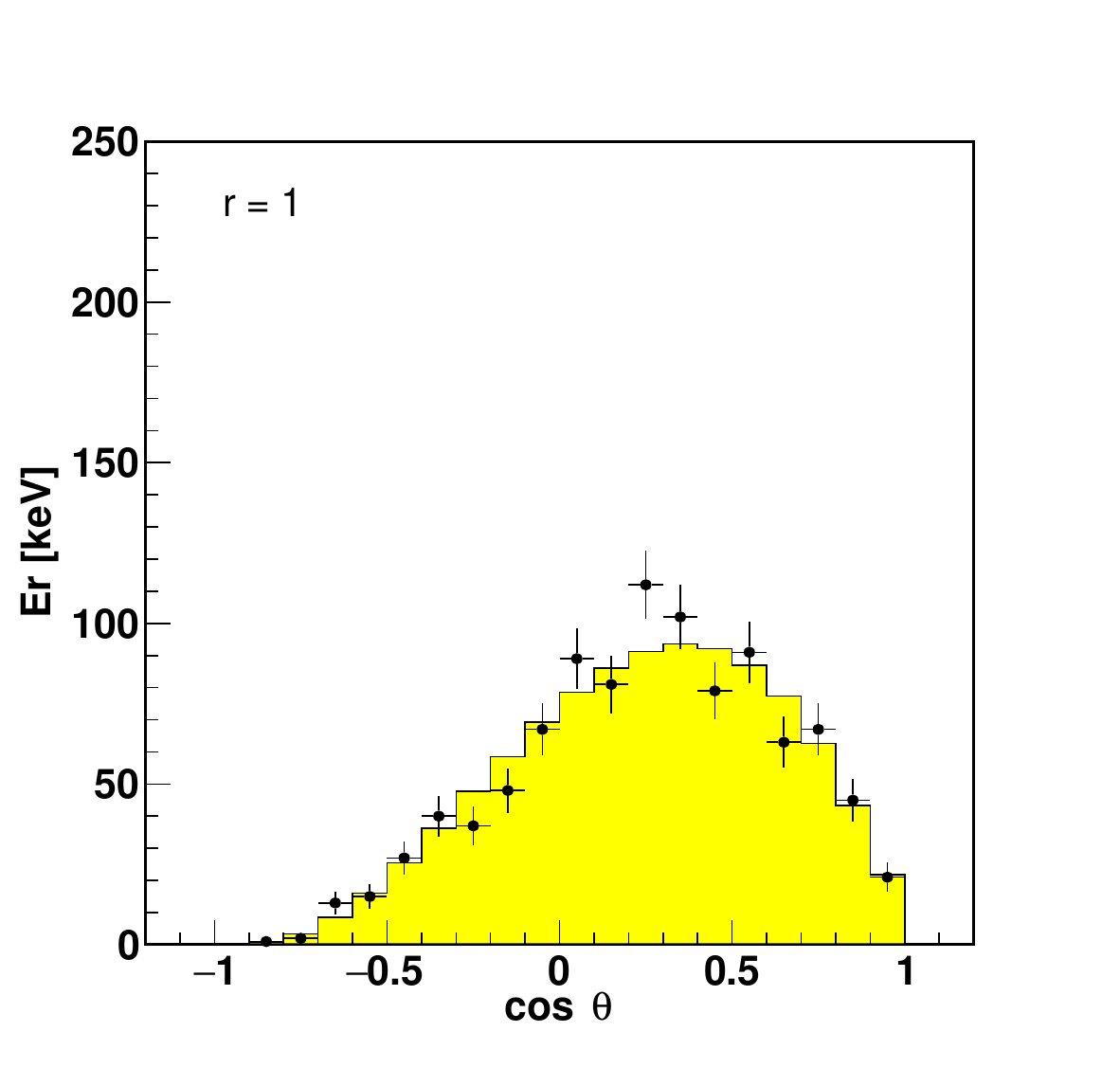}
 \caption{Histogram of $\cos\theta$ the for case of F as the target nucleon (light target), $m\chi=3m_A$, and the recoil energy cutoff  $E_R^{\mathrm{thr}} = 20$ keV. Yellow histograms correspond to the template (event number: $10^8$), and black points represent
 histograms for the pseudo-experiment (event number: $10^3$) together with statistically approximate error bars.}
 \label{fig:coshistF_ERnon0}
\end{figure}
\begin{figure}[h]
\includegraphics[keepaspectratio, scale=0.19,clip]{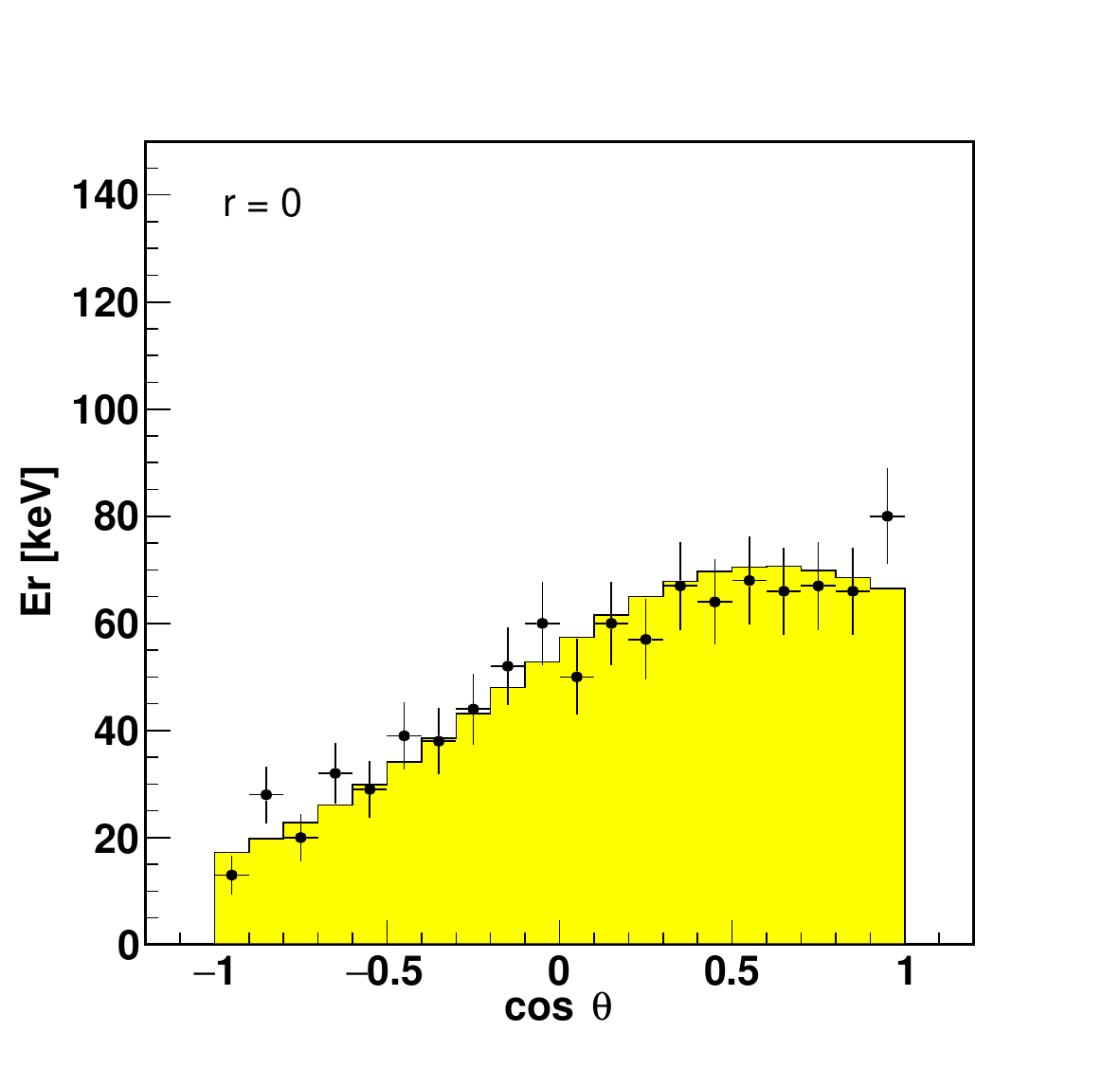}
\includegraphics[keepaspectratio, scale=0.19,clip]{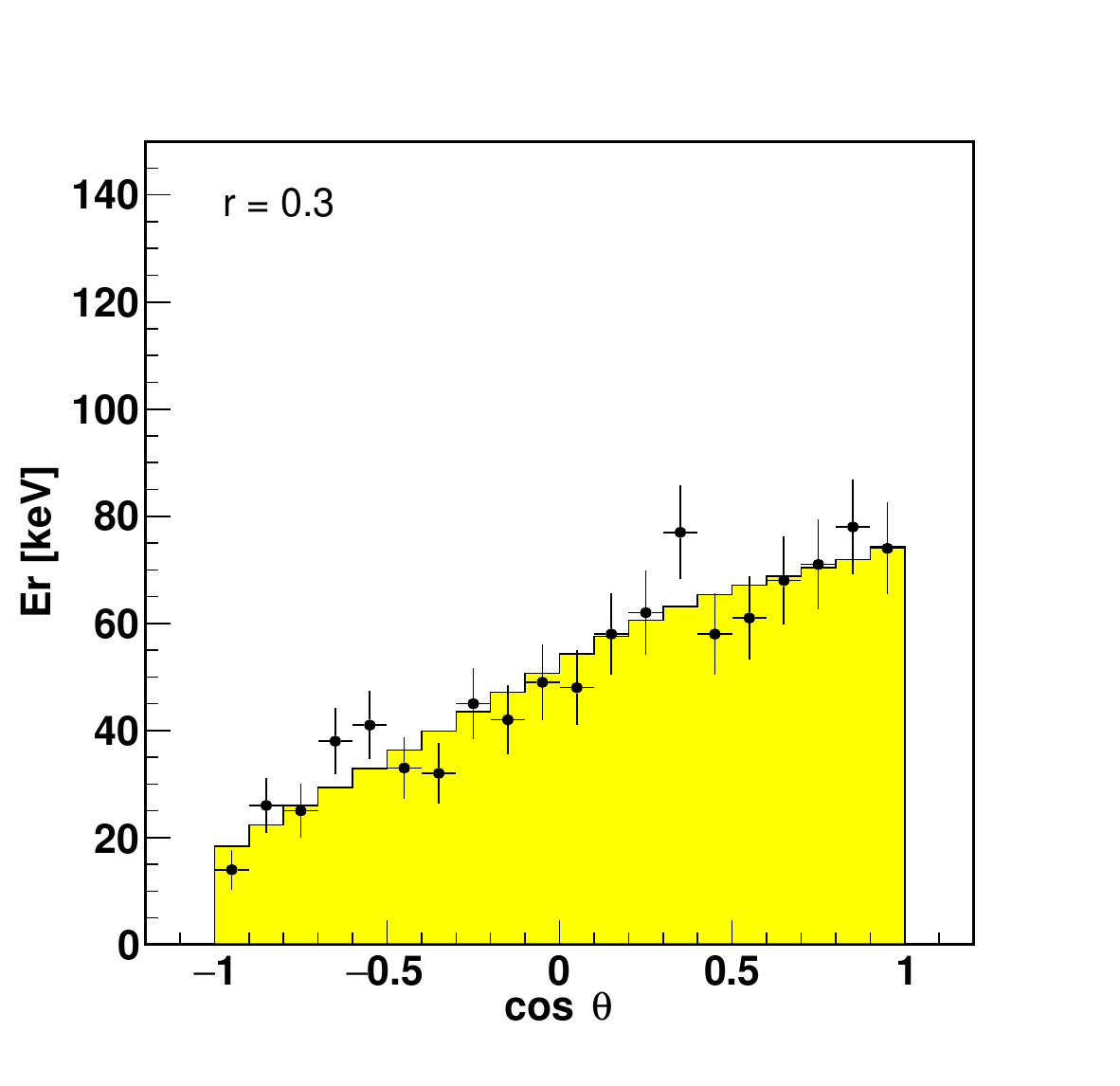}
\includegraphics[keepaspectratio, scale=0.19,clip]{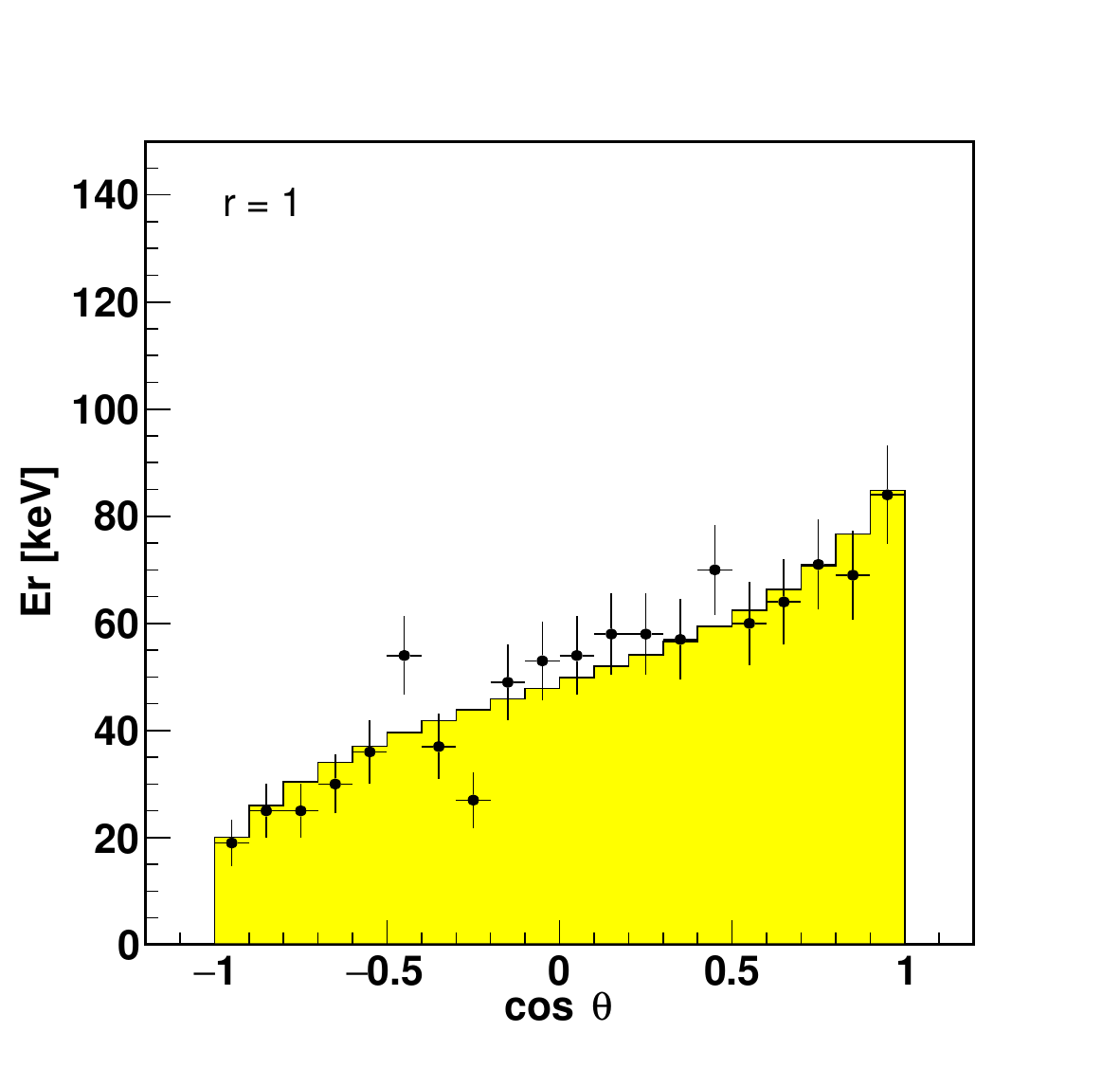}
 \caption{Histogram of $\cos\theta$ the for case of Ag as the target nucleon (heavy target), $m\chi=3m_A$, and the recoil energy cutoff  $E_R^{\mathrm{thr}} = 0$ keV. Yellow histograms correspond to the template (event number: $10^8$), and black points represent
 histograms for the pseudo-experiment (event number: $10^3$) together with statistically approximate error bars.}
 \label{fig:coshistAg_ER0}
\end{figure}
\begin{figure}[h]
 \centering
\includegraphics[keepaspectratio, scale=0.19,clip]{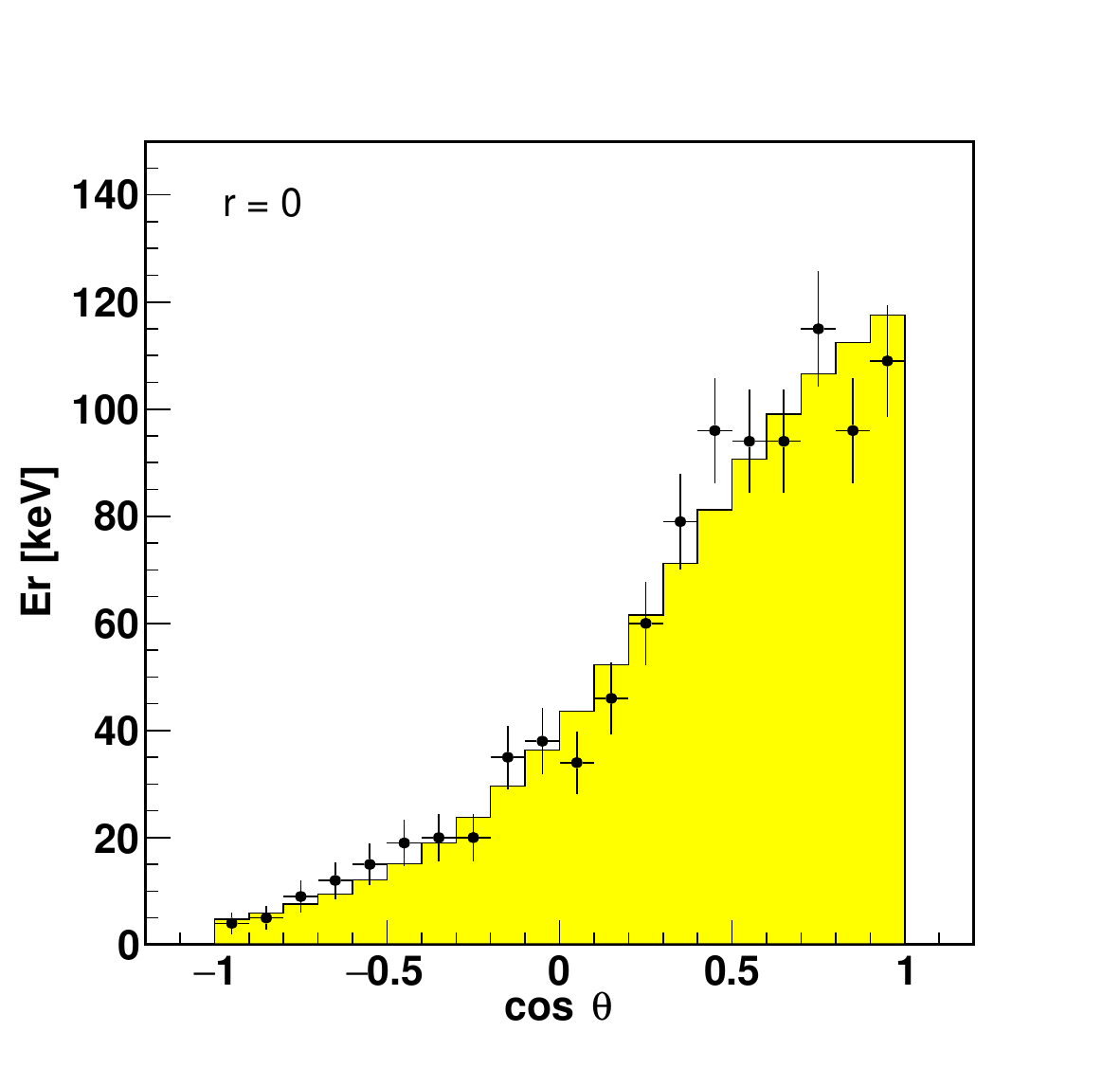}
\includegraphics[keepaspectratio, scale=0.19,clip]{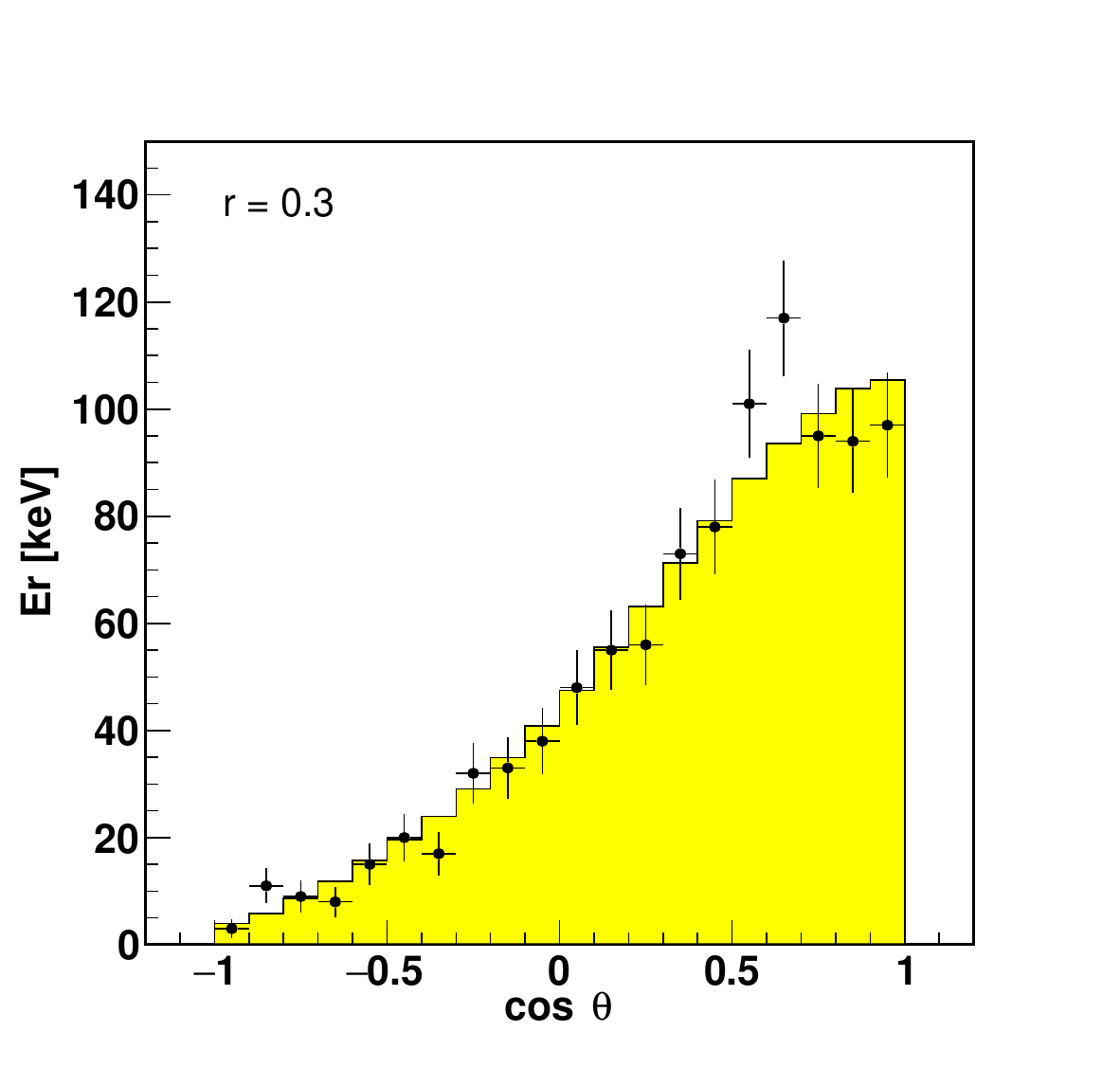}
\includegraphics[keepaspectratio, scale=0.19,clip]{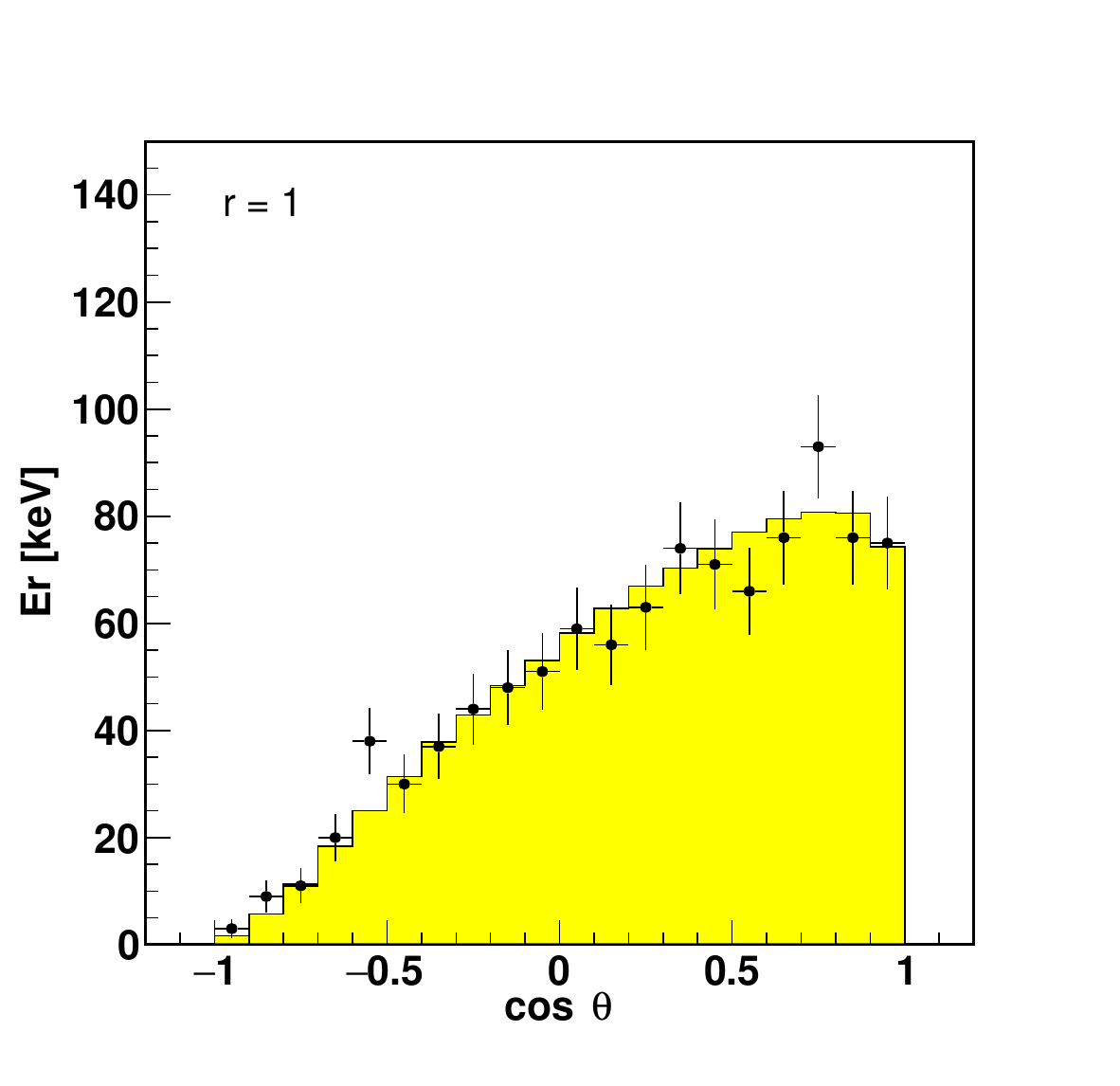}
 \caption{Histogram of $\cos\theta$ the for case of Ag as the target nucleon (heavy target), $m\chi=3m_A$, and the recoil energy cutoff  $E_R^{\mathrm{thr}} = 50$ keV. Yellow histograms correspond to the template (event number: $10^8$), and black points represent
 histograms for the pseudo-experiment (event number: $10^3$) together with statistically approximate error bars.}
 \label{fig:coshistAg_ERnon0}
\end{figure}

In Figure \ref{fig:coshistF_ER0}--\ref{fig:coshistAg_ERnon0}, histograms of the angle $\cos\theta$ are shown. 
Yellow histograms in the figures represent the template results. The corresponding pseudo-experiment results 
with statistically approximate error bars are shown in black. 
The template histograms are scaled by the number of events to the pseudo-experiment results for an easy comparison.
In Figure \ref{fig:coshistF_ER0}--\ref{fig:coshistF_ERnon0} the target is F, whereas the target 
in Figure \ref{fig:coshistAg_ER0}--\ref{fig:coshistAg_ERnon0} is Ag.
The threshold of the recoil energy is taken as $0$ keV (Figure \ref{fig:coshistF_ER0}), $20$ keV (Figure \ref{fig:coshistF_ERnon0}) and 
$0$ keV (Figure \ref{fig:coshistAg_ER0}), $50$ keV (Figure \ref{fig:coshistAg_ERnon0}). 
The non-realistic cases of $E_R^\mathrm{thr} = 0$ keV are presented for reference.

The anisotropic feature is seen in the angular distribution as a broad peak around  $\cos\theta\sim 0.4$ in the most distinctive
case of $r=1.0$, target F and $E_R^\mathrm{thr}=20$ keV.
If the velocity distribution is isotropic, the signal number is large in $\cos\theta \simeq 1$, whereas the signals tend to be scattered even in the  
$\cos\theta \ll 1$ region with strong anisotropy.
Choice of energy threshold $E_R^{\mathrm{thr}}$  is made in order to clearly demonstrate 
the anisotropy dependence of the histogram.
It is seen that isotropic distribution ($r=0$) makes the peak around $\cos\theta\sim 1$ in the four cases. It is interesting to point out that more clear peaks are seen with realistic energy threshold case than with $E_R^\mathrm{thr}=0$ case. This can be explained that the recoils to forward direction get higher energy than the recoils to the right angle.

If the forward/backward direction of nuclear recoil 
tracks are detected, a survey from $\cos\theta = -1 $ to $1$ can be completed as shown in the relevant figures. 
The anisotropy discrimination based on a perfect forward/backward detection will be discussed in the following section.
Most of the existing detectors, however, do not show perfect forward/backward detection and the absolute value $|\cos\theta|$ is adopted.

\section{Anisotropy Test}
\label{sec:rdetermination}
In this section, the number of events required to discriminate or give a constraint to the anisotropy of the velocity distribution is discussed. In all of the cases chi-squared test is adopted, however, in some sample points we have checked that the required event number is in the same order of magnitude to the analysis by likelihood approach.
\subsection{Chi-squared test using the energy-angular distribution}
\label{subsec:chi2_distribution}
\begin{figure}[t]
 \centering
 \vspace{-0.5cm}
 \includegraphics[keepaspectratio, scale=0.4]{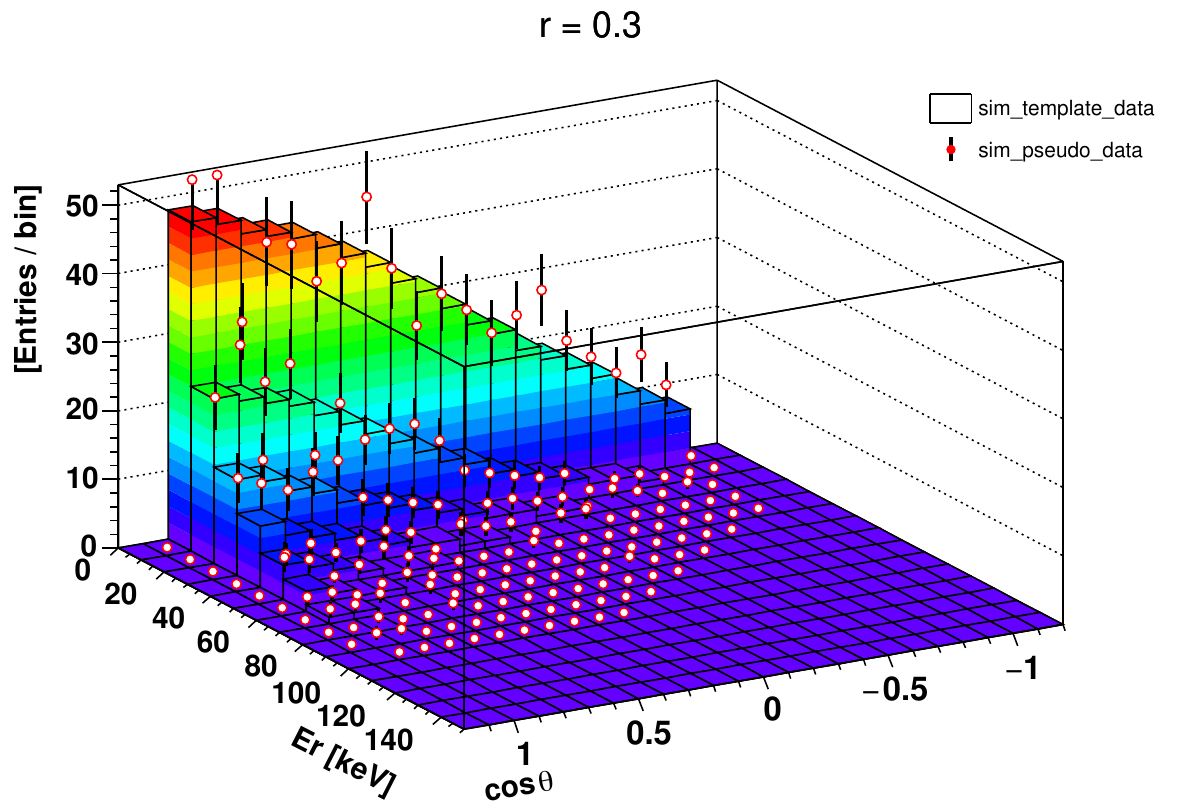}
 \caption{Three-dimensional plot of $r=0.3$ case in Figure \ref{fig:ERcos_F}. The $z$-axis represents event number for each bin. 
 The colored histogram and white dots correspond to template and pseudo-experiment cases, respectively.}
\label{fig:temppseudo_overlay}
\end{figure}
\begin{figure}[t]
 \centering
 \includegraphics[width=5cm,clip]{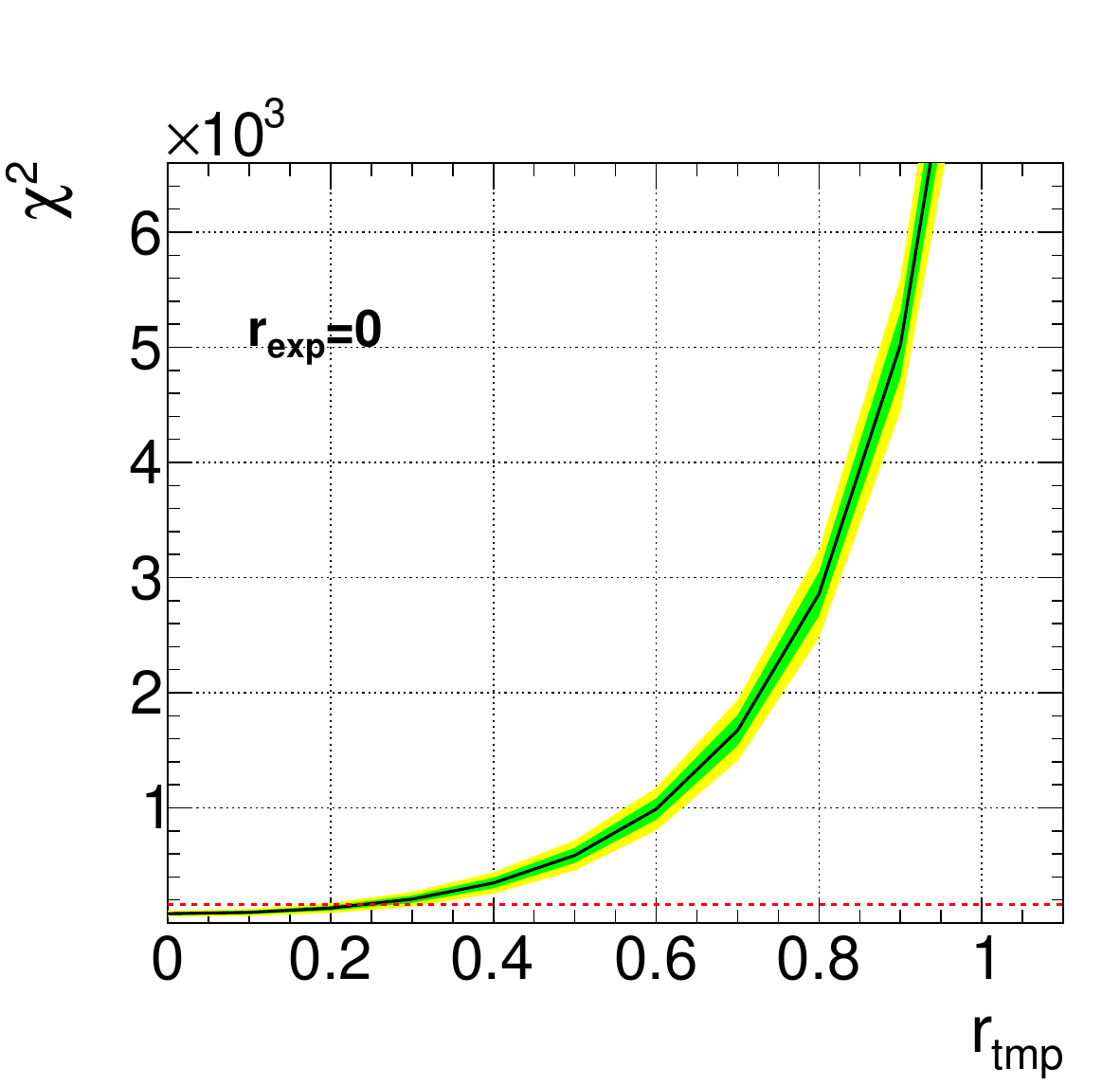}
  \includegraphics[width=5cm,clip]{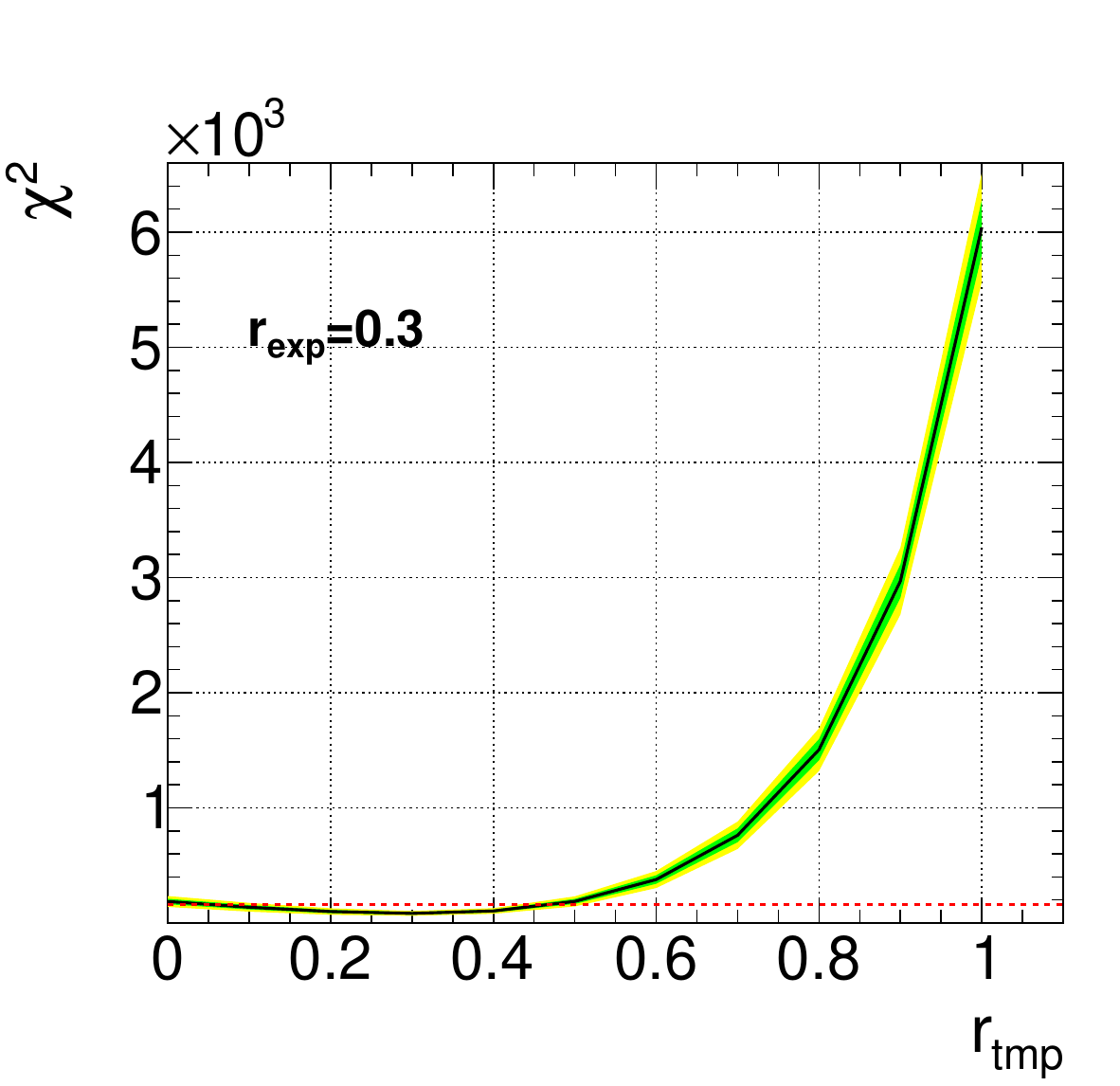}
 \caption{The chi-squared test of the energy-angular distributions, between the anisotropy of template $r_\mathrm{tmp}$ and that for the pseudo-experiment $r_\mathrm{exp}$. The target atom is F and the energy threshold $E_R^{\mathrm{thr}} = 20$ keV. In the pseudo-experiment 
the event number $6\times 10^3$. The red dashed line represents 90\% CL. Green and yellow green bands correspond to 68 \% and 95 \% C.L., respectively.}
\label{fig:chi2_FERnon0}
\end{figure}
\begin{figure}[t]
 \centering
 \includegraphics[width=5cm,clip]{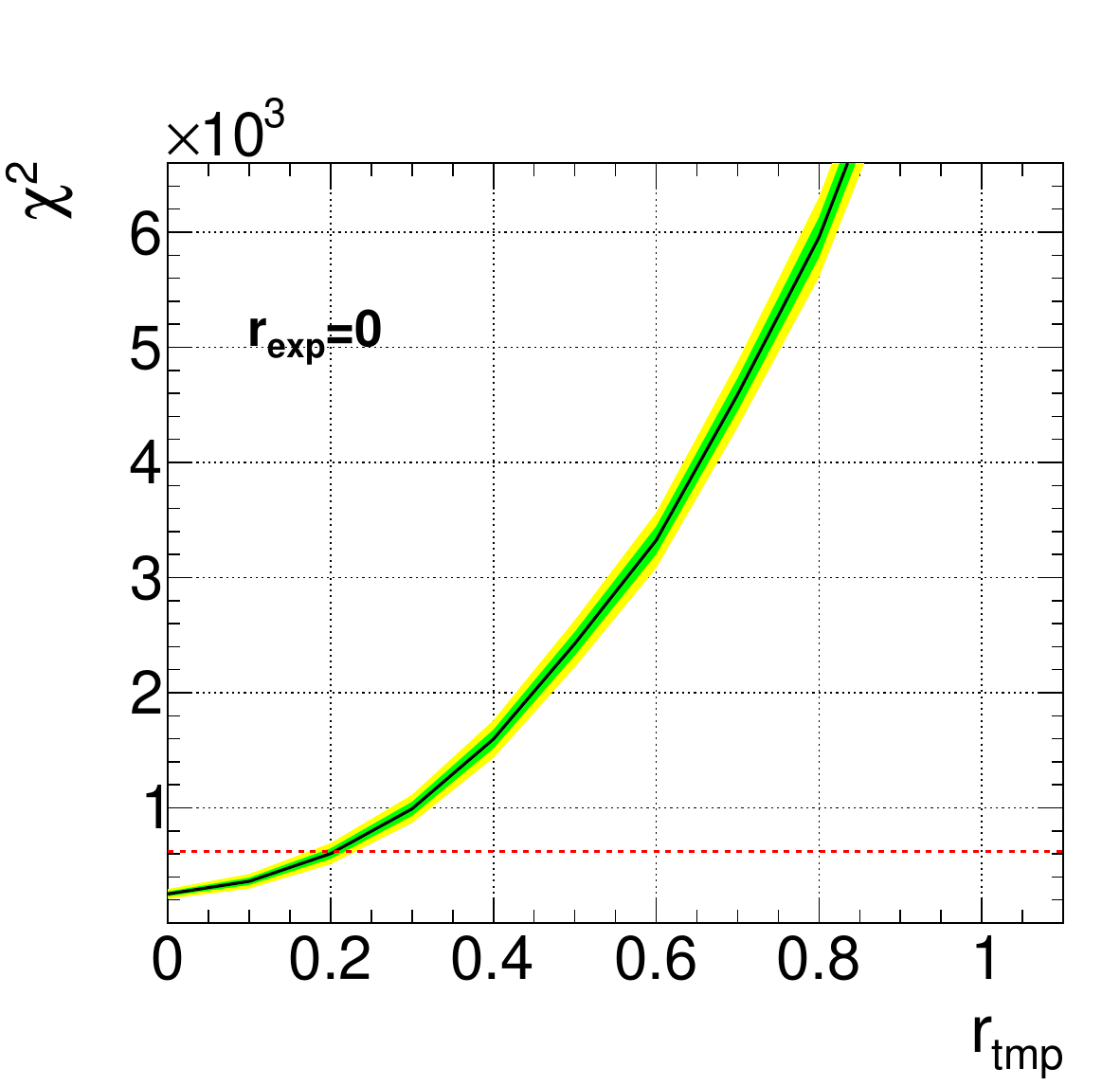}
  \includegraphics[width=5cm,clip]{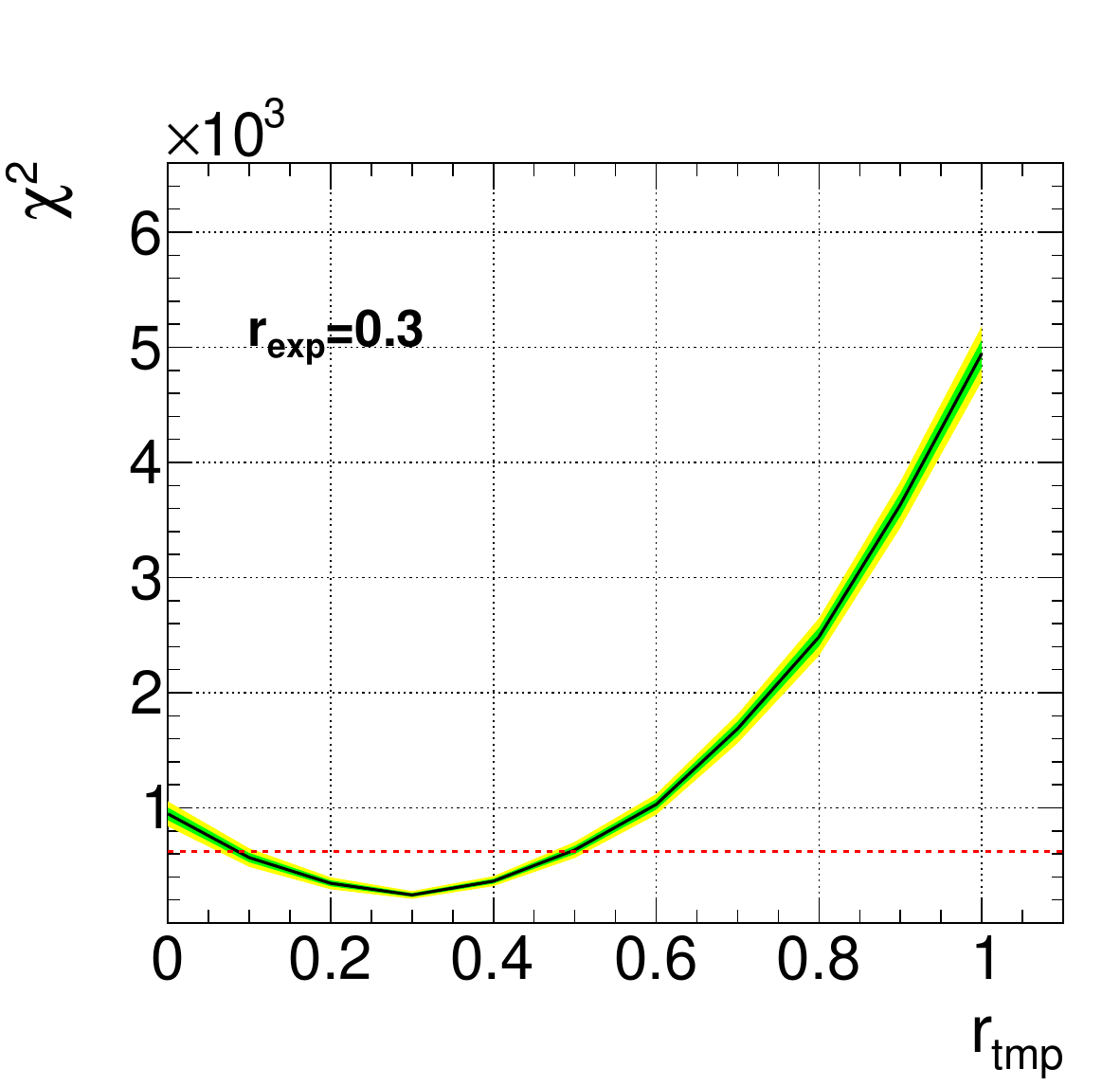}
 \caption{The chi-squared test of the energy-angular distributions, between the anisotropy of template $r_\mathrm{tmp}$ and that for the pseudo-experiment $r_\mathrm{exp}$. The target atom is Ag and the energy threshold $E_R^{\mathrm{thr}} = 50$ keV. In the pseudo-experiment 
the event number $6\times 10^4$. The red dashed line represents 90\% CL. Green and yellow green bands correspond to 68 \% and 95 \% C.L., respectively.}
\label{fig:chi2_AgERnon0}
\end{figure}

Three-dimensional plots for the case $r=0.3$ case (as seen in Figure \ref{fig:ERcos_F}) of the template data
and the pseudo-experimental data are shown in Figure \ref{fig:temppseudo_overlay}.
Here, both of the $r$ value for the template ($r_\mathrm{tmp}$) and the pseudo-experiment ($r_\mathrm{exp}$) are set to 0.3 and it is seen that the two plots agree within the statistical fluctuation.
The $E_R-\cos{\theta}$ plane is divided into 10 ($E_R$) $\times$ 20 ($\cos{\theta}$)  bins, and each cuboid and white dot represent
the event number for the scaled template and pseudo-experiment, respectively. 
The chi-square value between the template and pseudo-experiment for each set of $r_\mathrm{tmp}$ and $r_\mathrm{exp}$ is calculated.
For reference, chi-square value between the template and the pseudo experiment are shown
in Figure \ref{fig:chi2_FERnon0} for $E_R^\mathrm{thr}=20$ keV (a realistic case).
Here, the target nuclei is fluorine. 
Each figure show the chi-squre values for a given $r_\mathrm{exp}$ with the $r_\mathrm{tmp}$ as the horizontal axis.
Same plots for the silver target with $E_R^\mathrm{thr}=50$ keV are shown in Figure \ref{fig:chi2_AgERnon0}. 
 Red dashed lines in the figures correspond to a 90\% CL. 
Light green and yellow region in the figures correspond to 1$\sigma$ and 2$\sigma$ statistical fluctuation bands estimated by 100 trials of the pseudo-experiments.


As stated above, the energy threshold is a factor that clearly characterizes the anisotropy 
as well as the number of signals in the experiment. 
Most of the events are concentrated in the low energy region $E_R\sim 0$ keV, and imposing the energy threshold difference 
between the distributions can be clear. 
In Figures \ref{fig:chi2_FERnon0} and \ref{fig:chi2_AgERnon0}, the energy threshold is set 
at $E_R^\mathrm{thr} = 20$ keV for the target F and $50$ keV for the target Ag, respectively.
Anisotropic halo models ($r_\mathrm{tmp}>0$) are tested by the results in Figure \ref{fig:chi2_FERnon0} and \ref{fig:chi2_AgERnon0}. For each $r_\mathrm{exp}$ value, 
which we assume as the real $r$ value, the $r_\mathrm{tmp}$ models which show chi-square values larger than the 90\% CL can be rejected.  
The most interesting case tested by this experiments would by discriminating isotropic halo model of $r=0$ from anisotropic halo model of $r=0.2$--$0.3$  which is 
suggested by N-body simulations in previous studies \cite{LNAT}.
Figure \ref{fig:chi2_FERnon0} shows that the experimental data of $r_\mathrm{exp}=0.3$ with $6\times 10^3$ fluorine recoil events can reject $r=0$ model at 90\% C.L. 
In the case of the silver target, same rejection is obtained with a statistics of 6 $\times 10^4$ events.
In particular, for the case of the heavy target like silver, the shape of the energy-angular distribution is largely distorted 
by the effect of the form factor in the region of 40 keV or more as clearly illustrated as a bump, and this may be used to distinguish 
distributions with different ratios from each other. 
However, for the heavy target, a greater event 
number is needed to compare the result of the pseudo-experiment result with those of the template case 
as opposed to the case of the light target. 

\subsection{The chi-squared test using the angular histogram}
\label{subsec:chi2_histogram}
\begin{figure}[t]
 \centering
 \includegraphics[width=5cm,clip]{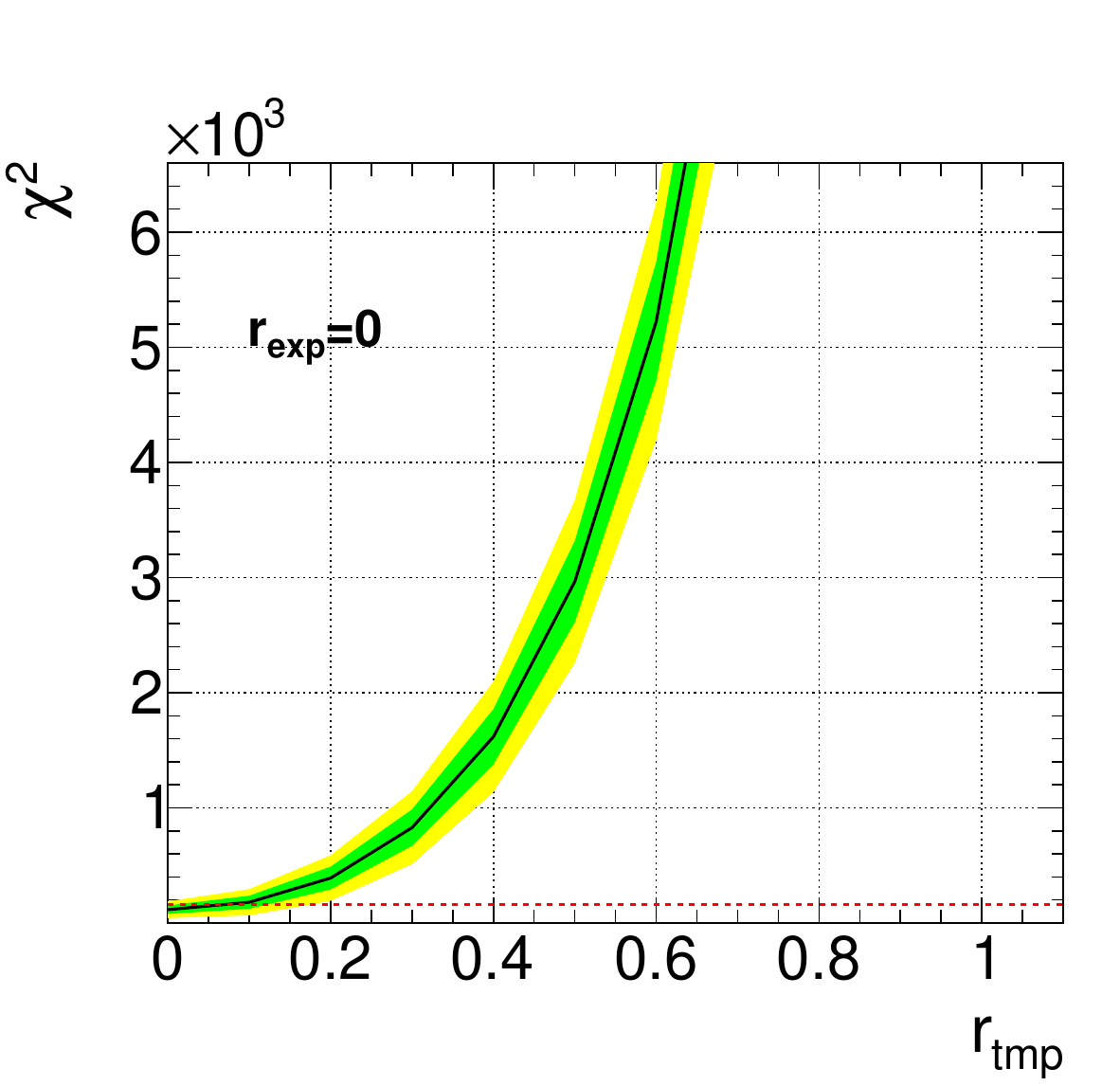}
 \includegraphics[width=5cm,clip]{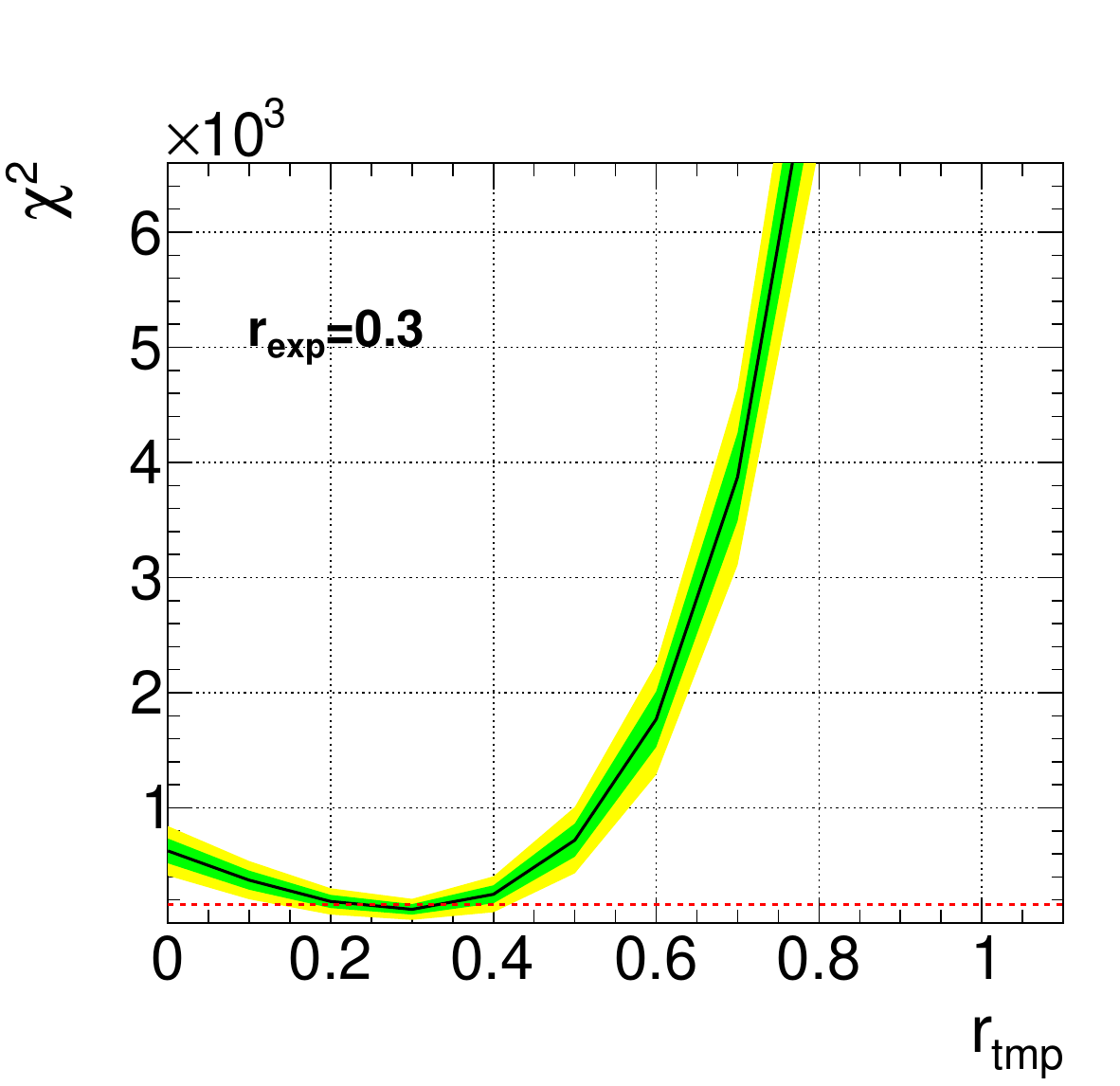}
 \caption{The chi-squared test of the angular histograms, between the anisotropy of template $r_\mathrm{tmp}$ and that for the pseudo-experiment $r_\mathrm{exp}$. The target atom is F and the energy threshold $E_R^{\mathrm{thr}} = 20$ keV. In the pseudo-experiment 
the event number $5\times 10^3$. The red dashed line represents 90\% CL. Green and yellow green bands correspond to 68 \% and 95 \% C.L., respectively.}
\label{fig:1D_chi2_FERnon0}
\end{figure}

\begin{figure}[t]
 \centering
 \includegraphics[width=5cm,clip]{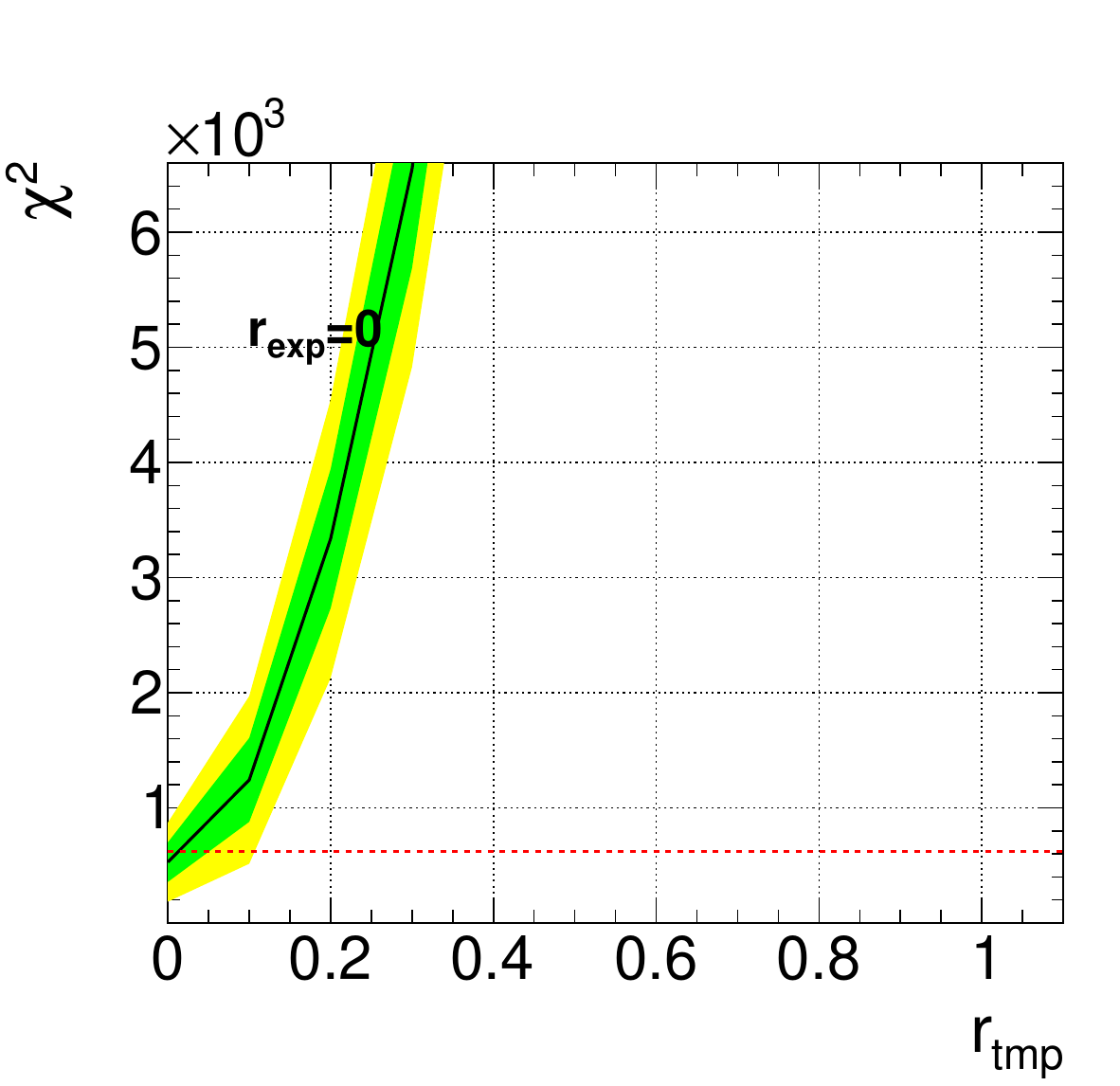}
  \includegraphics[width=5cm,clip]{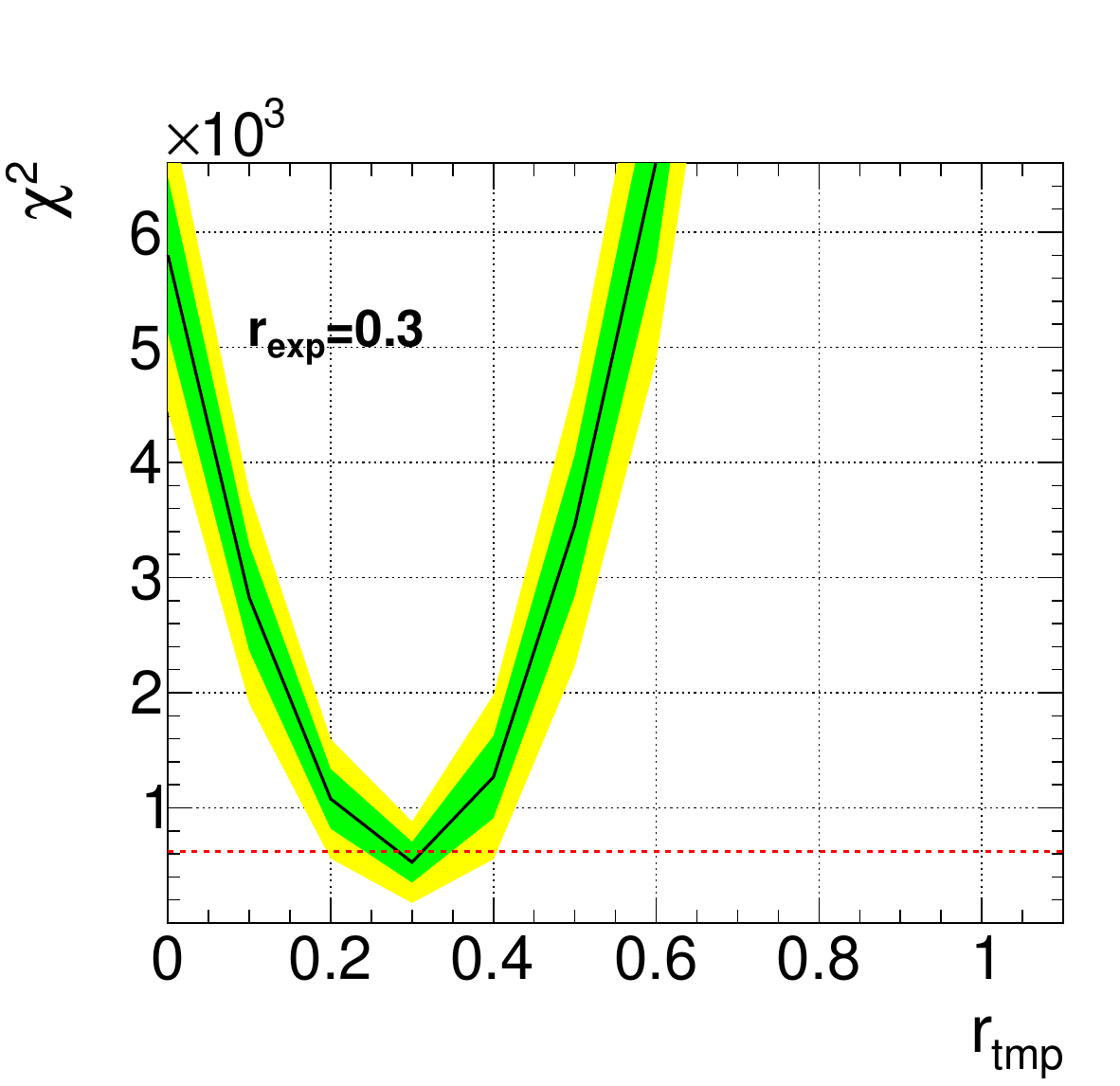}
 \caption{The chi-squared test of the angular histograms, between the anisotropy of template $r_\mathrm{tmp}$ and that for the pseudo-experiment $r_\mathrm{exp}$. The target atom is Ag and the energy threshold $E_R^{\mathrm{thr}} = 50$ keV. In the pseudo-experiment 
the event number $2\times 10^4$. The red dashed line represents 90\% CL. Yellow green and green bands correspond to 68 \% and 95 \% C.L., respectively.}
\label{fig:1D_chi2_AgERnon0}
\end{figure}

Some directional detectors do not have  very good energy resolution and it is more realistic to 
use only the angular histograms making use of sufficient angular resolutions. 
With angular histograms shown in Figure \ref{fig:coshistF_ER0}--\ref{fig:coshistAg_ERnon0}, we can calculate chi-square
between the template data and pseudo-experimental data. 
Range $-1\leq \cos{\theta} \leq 1$ is divided into 20 bins and chi square value is calculated.
In Figure \ref{fig:1D_chi2_FERnon0} and Figure \ref{fig:1D_chi2_AgERnon0},
results of the chi-squared test for target F and Ag are shown, respectively. The energy thresholds are set at $E_R^\mathrm{thr} = 20$ keV in Figure \ref{fig:1D_chi2_FERnon0}
and $50$ keV in Figure \ref{fig:1D_chi2_AgERnon0}. 
In the figures, red dashed lines correspond to a 90 \% confidence level (CL), 
and light green and yellow region in the figures represent 68\% and 95\% confidence intervals, respectively. 
They correspond to 1$\sigma$ and 2$\sigma$ regions of a 100 times trials to produce the data sets.

For target F, $\sim 5\times 10^3$ events are required
to exclude r = 0 for the r = 0.3 case at 90\%CL, while for target Ag, $2\times 10^4$ events are required. 
Note that for the heavy target (Ag), a greater event 
number is needed to compare the result of the pseudo-experimental result with those of the template case 
as opposed to the case of the light target (F), since large number of events are lost due to the effect of nuclear form factor.

It is interesting to point out that
the discrimination can be possible with a smaller event number  than that of 
the energy-angular distribution case. It is because an event number for each bin of the chi-squared test in one dimensional
analysis tends to be large compared to two dimensional analysis. 
From these results, two possible approach can be proposed to 
discriminate this particular halo model under the condition that the dark matter mass is already known by accelerator experiments or astrophysical observations. One is to set an appropriate  analytical energy threshold and only use the angular histogram with data taken by a detector with sufficient energy threshold. This approach can be used to test other models and also to pin-down the dark mater mass with a precise energy-angular distributions. Another approach is to use a threshold-type detector without energy resolution with    
an appropriate energy threshold. 
Although this approach is very much model independent in terms of the particle physics and astrophysics, 
the cost would be lower than the former approach.  


\subsection{Constraint for the anisotropy in a case dark matter mass is not knowm}
\label{subsec:mdm_r_likelihood}
So far we have supposed that dark matter mass is known by collider experiment or astrophysical observation,
and fixed the dark matter mass in analysis. Although this is a case where the direction-sensitive method can be used effectively for the anisotropy study, it might not happen.
In this subsection, we consider a case that the dark matter mass is not known.
We estimate the parameters of the dark matter mass and the anisotropy by performing a binned likelihood analysis.
A likelihood function is defined by
\begin{equation}
L=\prod_{i=1}^{N_{\rm{ene}}}\prod_{j=1}^{N_{\rm{ang}}} P(M_{ij}|\bar{M}_{ij}) \ ,
\end{equation}
where $P(M_{ij}|\bar{M}_{ij})$ is the Poisson 
distribution function with the observed number of events $M_{ij}$ and the expected number of events $\bar{M}_{ij}$. The index of $i$ and 
$j$ represent the $i$-th energy bin and $j$-th angular bin.
The expected number of events $\bar{M}_{ij}$ was obtained using the template mentioned in Section \ref{sec:detailsoftheanalysis}.
We calculated the likelihood function employing MULTINEST\cite{Feroz:2007kg} in order to sample the likelihood function within the parameter space assuming flat priors and obtained posterior probability distributions. Also we assumed three detector configurations; with only energy, only direction, and both energy and direction information.

\begin{figure}[ht]
 \centering
 \includegraphics[keepaspectratio, scale=0.19,clip]{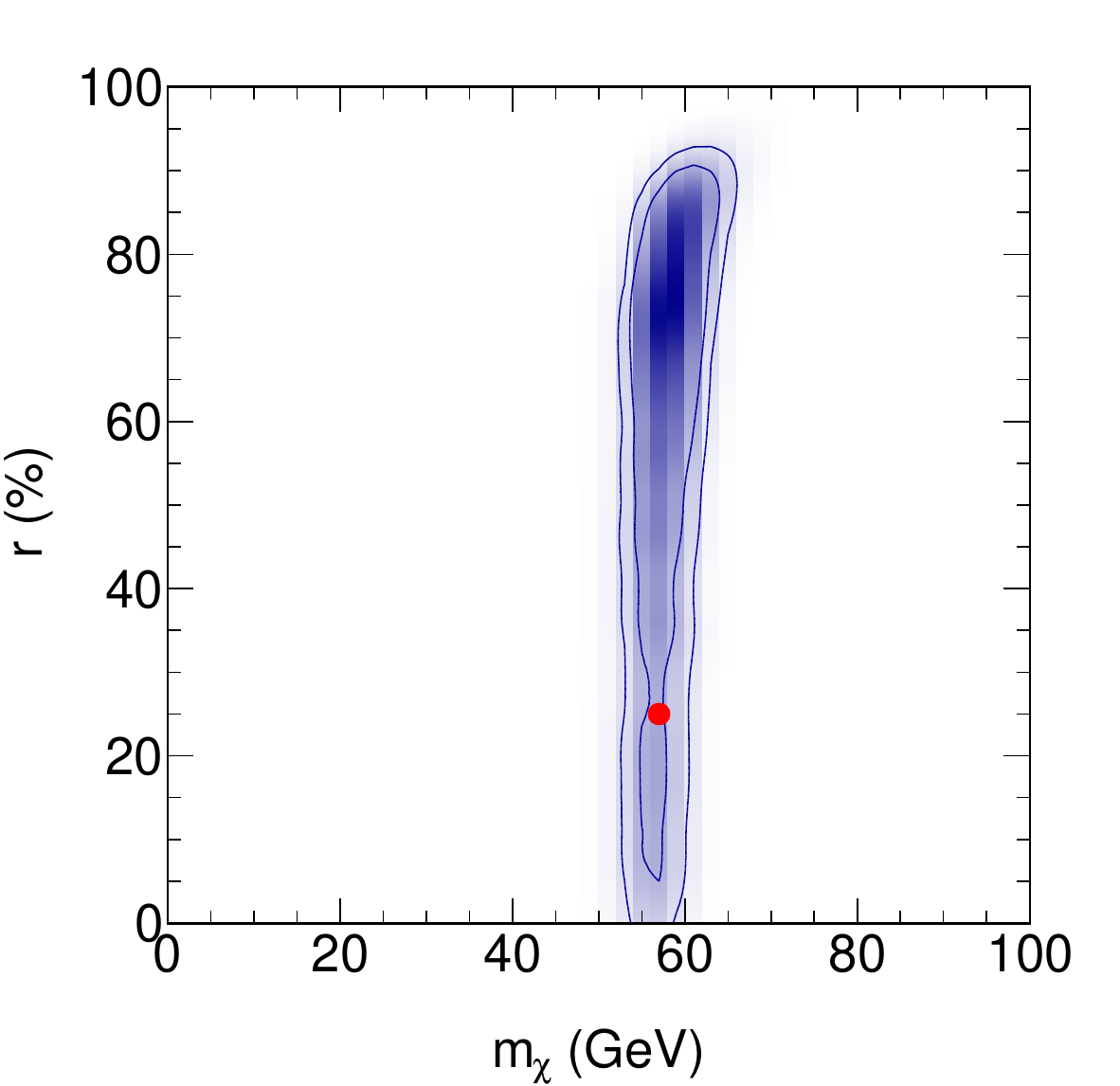}
\includegraphics[keepaspectratio, scale=0.19,clip]{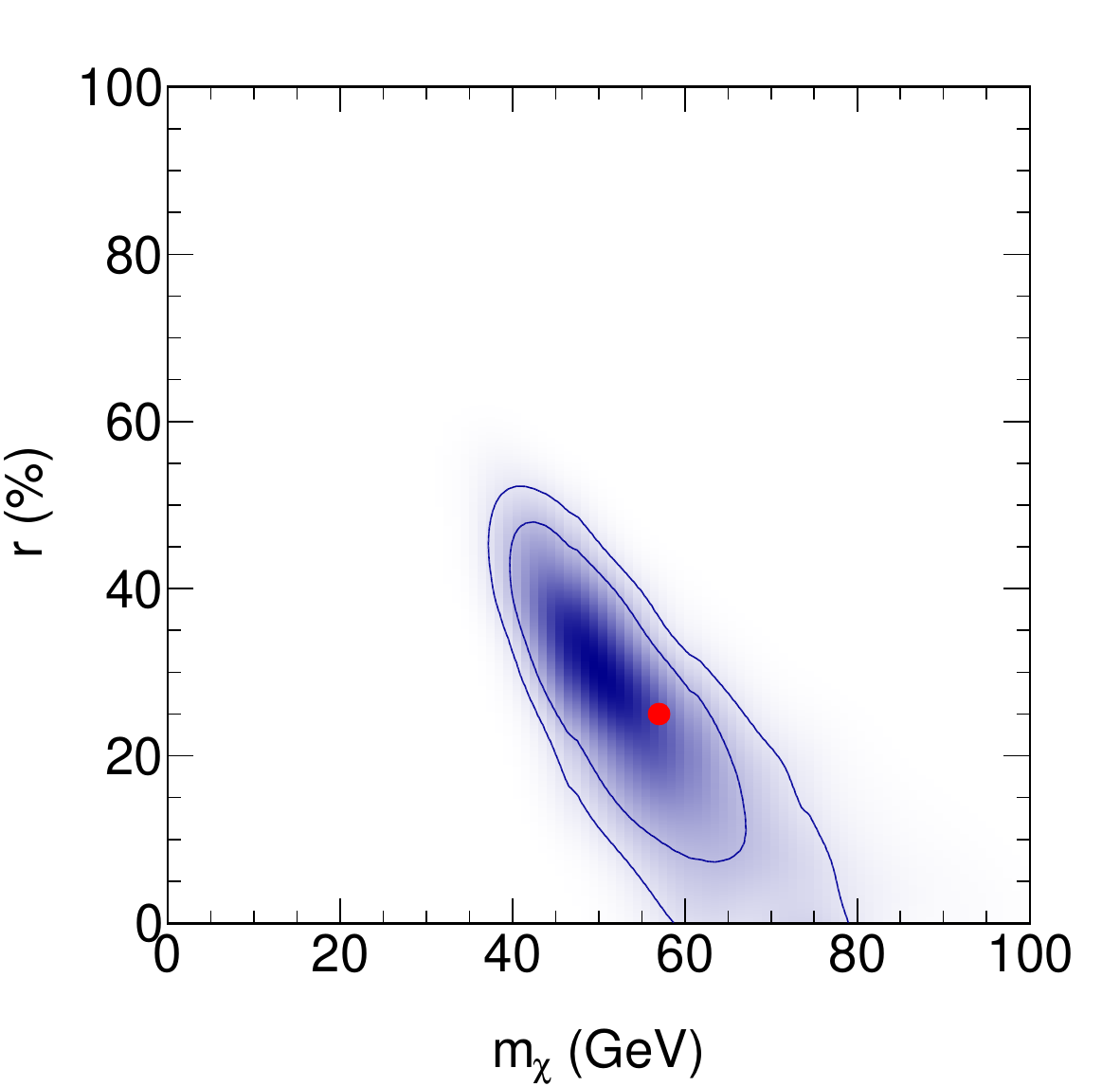}
\includegraphics[keepaspectratio, scale=0.19,clip]{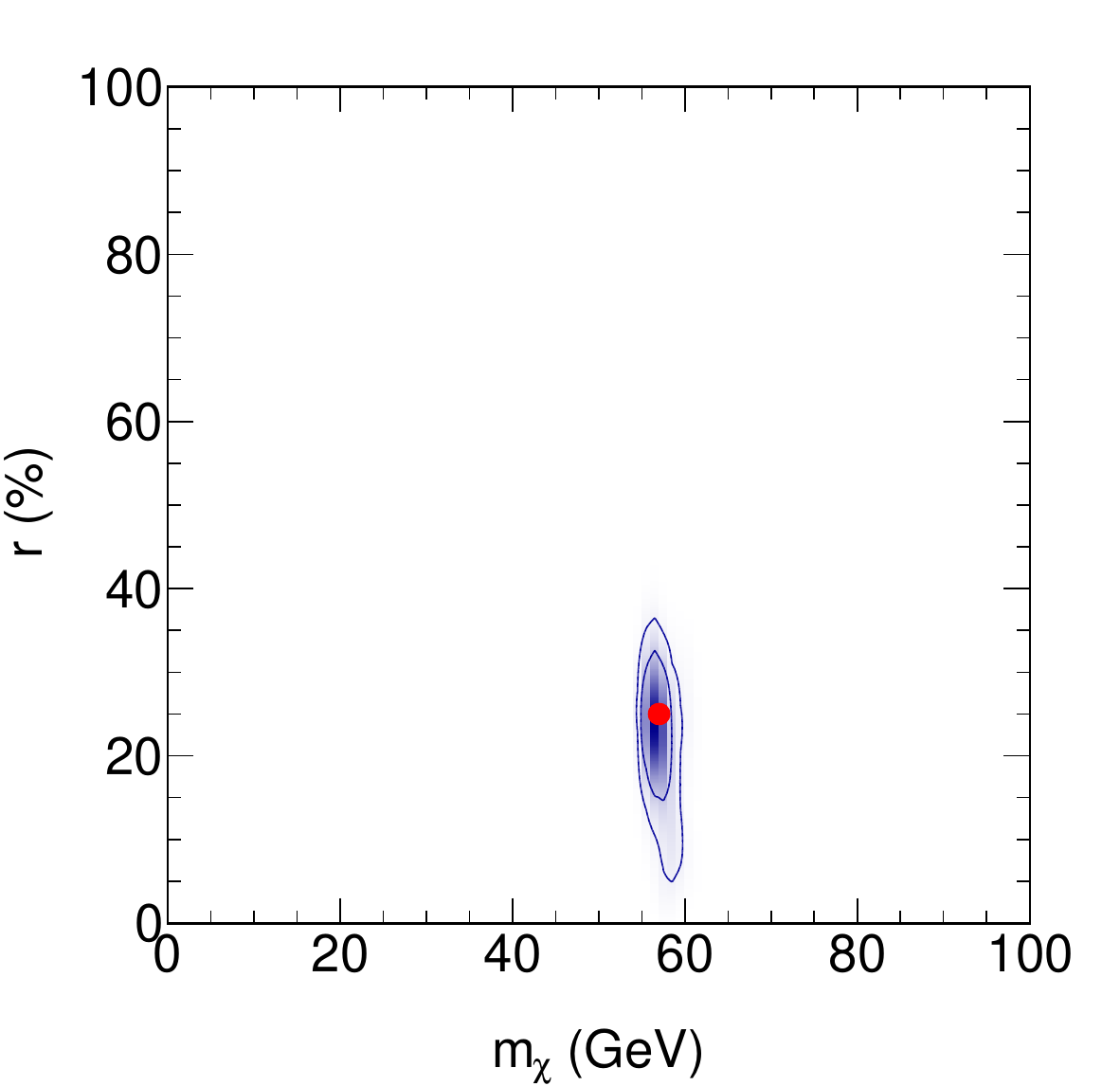}
 \caption{The 2D posterior probability distributions in the dark matter mass and anisotropy space for target F. Red points indicate  the input parameters in simulation. Inner and outer contours show 68\% and 90\% C.L, respectively. 
 Left: Only data of recoil energy $E_R$ is used. Center: Only data of scattering angle $\cos{\theta}$ is used. Right: Both recoil energy and scattering angle are used.}
 \label{fig:likelihoodF}
\end{figure}
\begin{figure}[ht]
 \centering
 \includegraphics[keepaspectratio, scale=0.19,clip]{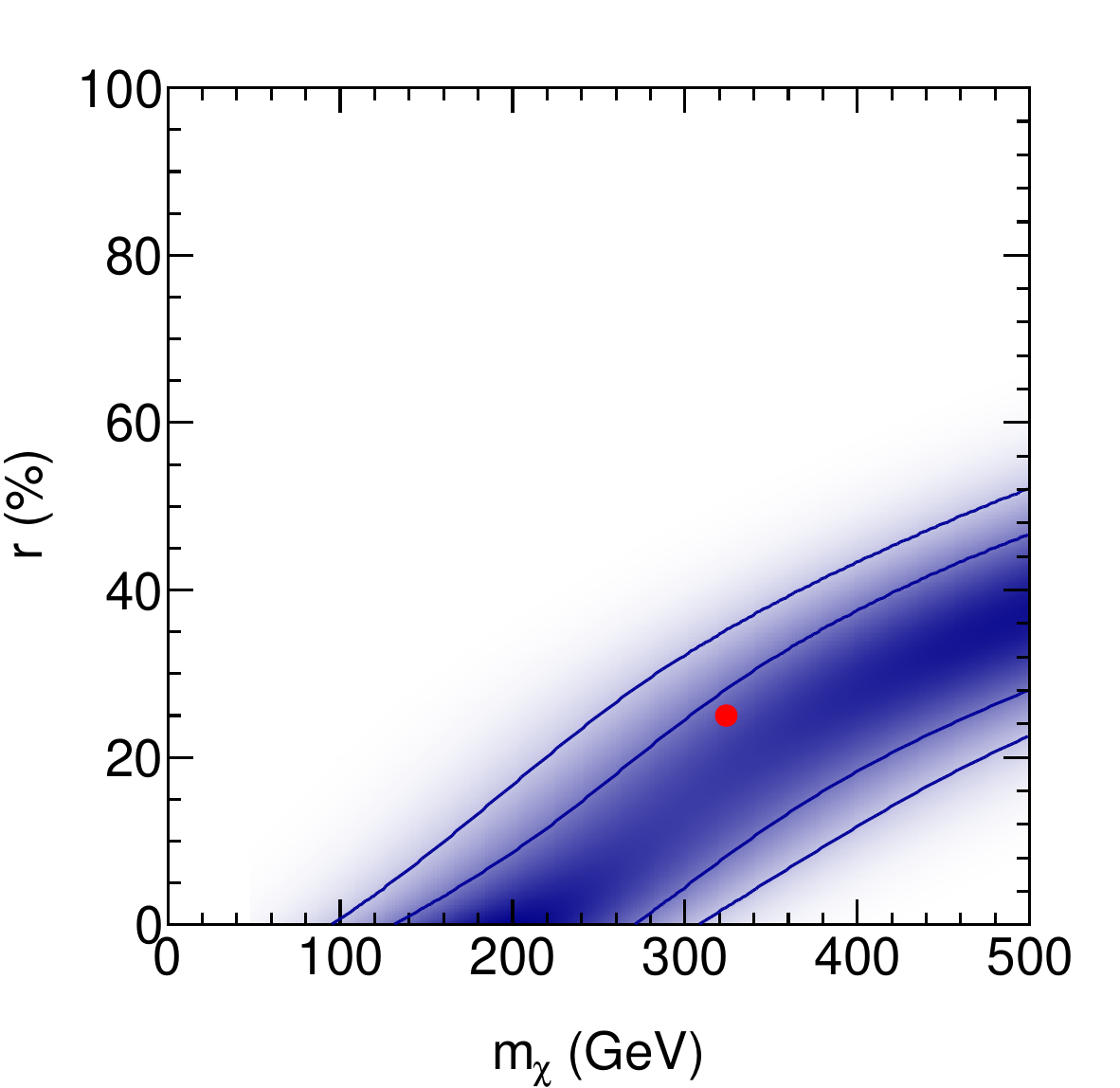}
\includegraphics[keepaspectratio, scale=0.19,clip]{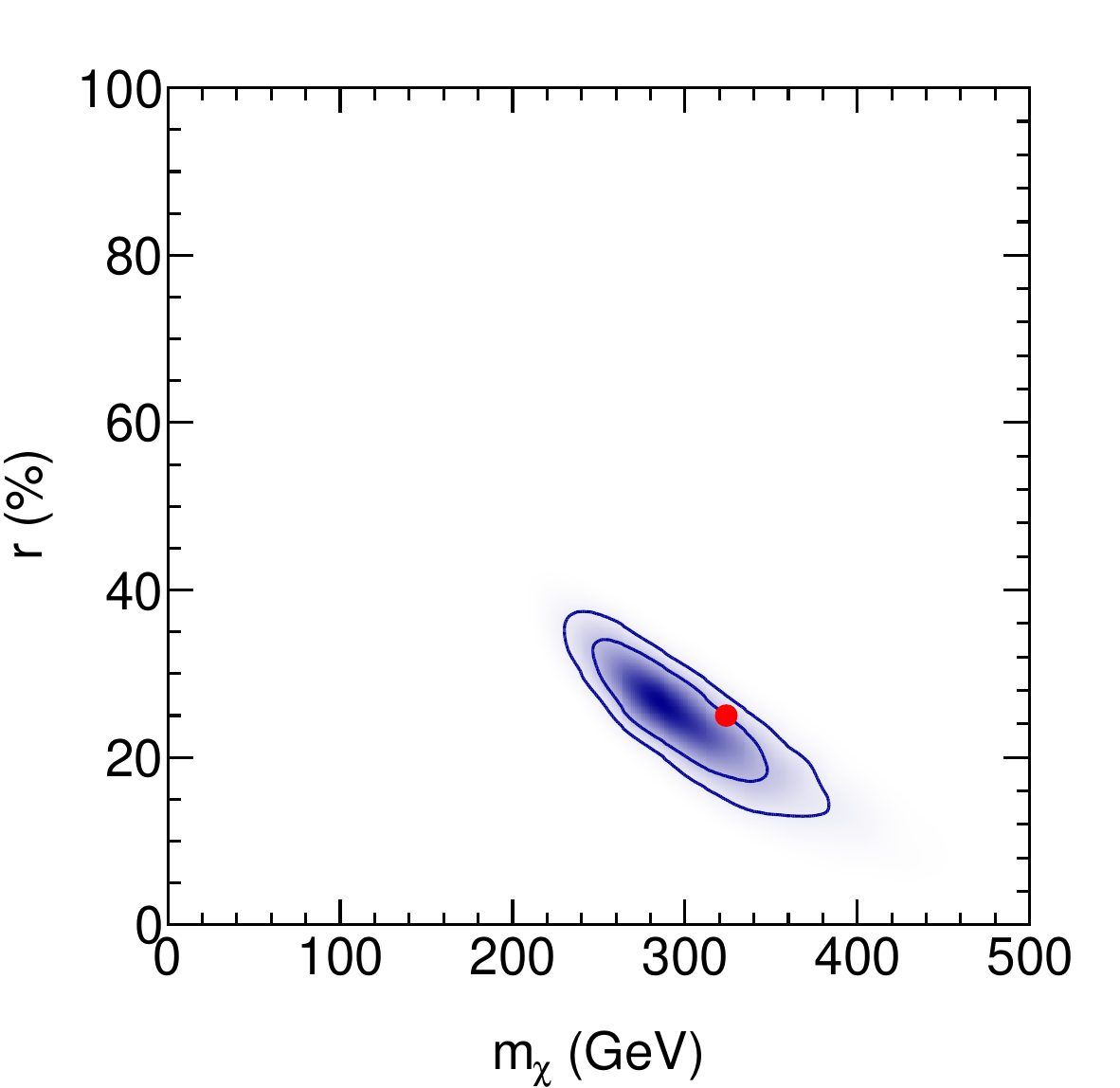}
\includegraphics[keepaspectratio, scale=0.19,clip]{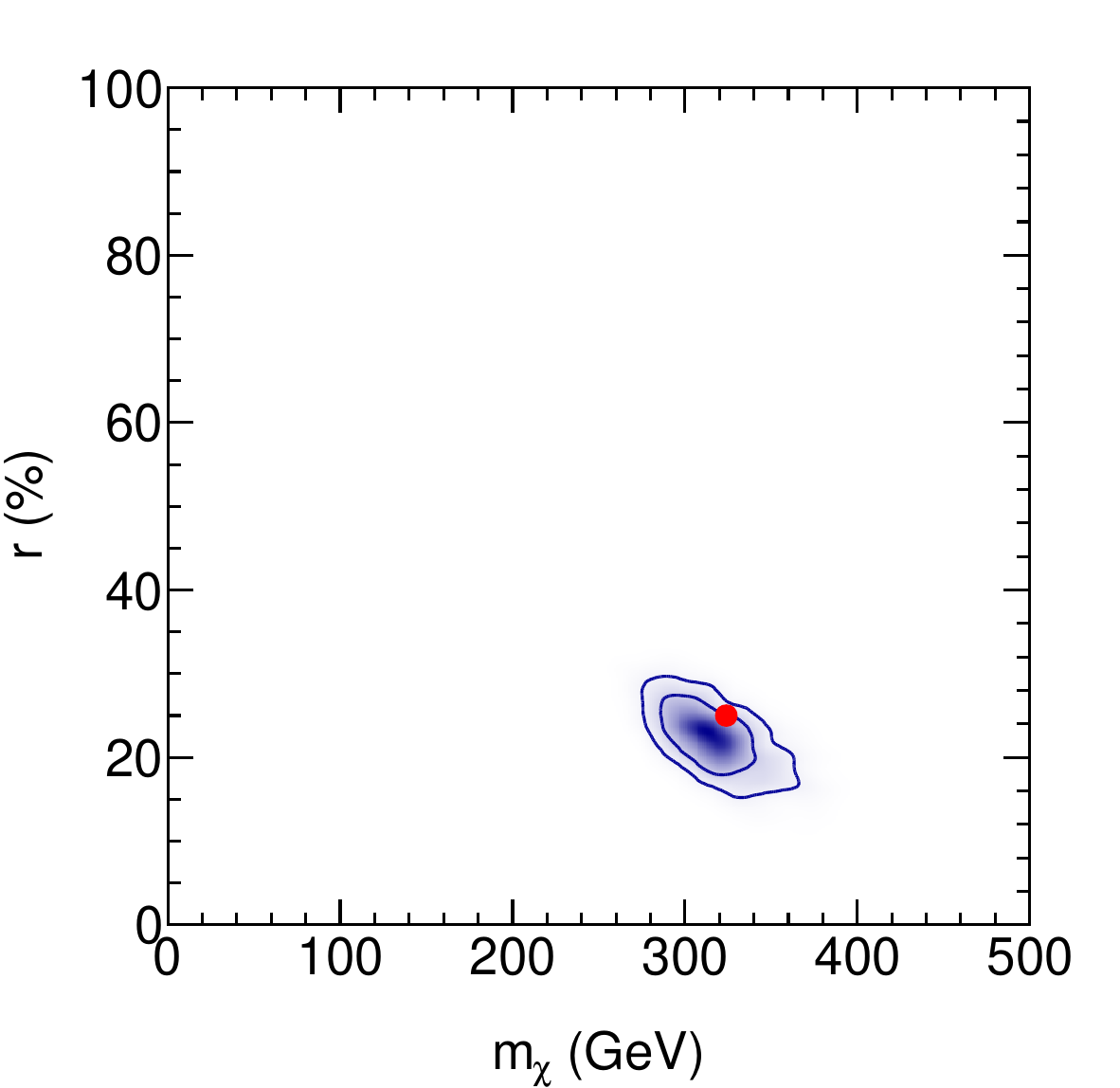}
 \caption{The 2D posterior probability distributions in the dark matter mass and anisotropy space for target Ag. Red points indicate the input parameters in simulation. Inner and outer contours show 68\% and 90\% C.L, respectively. 
 Left: Only data of recoil energy $E_R$ is used. Center: Only data of scattering angle $\cos{\theta}$ is used. Right: Both recoil energy and scattering angle are used.}
 \label{fig:likelihoodAg}
\end{figure}

In Figure \ref{fig:likelihoodF} and \ref{fig:likelihoodAg}, results of three analysis about for F target and Ag target case are shown, respectively. Red points correspond supposed parameters in simulation, i,e, $m_\chi=$57 GeV (F), $m_\chi=$324 GeV (Ag) and $r=0.25$. Inner and outer circles present 95\%, 68\% C.L, respectively.  By estimations using only recoil energy data, especially it is difficult to specify anisotropy, while uncertainly is reduced in analysis using only directional information. Uncertainty of dark matter mass and recoil energy is smaller in middle figures than that in right figures even if event numbers of pseudo experiment are same in the figures. Note that in Subsection \ref{subsec:chi2_distribution} and \ref{subsec:chi2_histogram}, required event number for analysis in energy-angular distribution is smaller than that in angular histogram. On the other hand, to give constraint for both dark matter mass and anisotropy at the same time, analysis using both recoil energy and scattering angle is most reasonable. Analysis using only recoil energy $E_R$ can be interpreted as analysis in the ordinary direct detection. In other words, Figure \ref{fig:likelihoodF} and \ref{fig:likelihoodAg} shows how much the directional detector can improve the parameter determination.


\section{Conclusion}
\label{sec:conclusion}
The directional detection of dark matter is a next-generation experiment. In this study,
 the potential of the directional detector to discriminate the 
anisotropic distribution of the dark matter velocity from the isotropic distribution is investigated.
The results of a Monte-Carlo simulation of the elastic 
scattering between a WIMP and a nucleon are statistically analyzed. 
Two cases are assumed depending on the energy resolution of 
the detector. If the energy resolution is mediocre, an angular histogram of nuclear recoils is obtained. 
If the velocity distribution is fully isotropic, the peak of the histogram is at $\cos{\theta}\sim 1$, 
while event number in the $\cos{\theta} \ll 1$ region increases as 
the velocity distribution has a more anisotropic component.
With good energy resolution, the energy-angular distribution may be analyzed. 
By means of a complete distribution
generated by a Monte-Carlo simulation, anisotropy can be presumed. 
Particularly, for the case of anisotropy $r=0.3$,
the completely isotropic case $r=0$ can be excluded at the 90 \% CL with $O(10^4)$ events 
for the energy-angular distribution case. 
It is also interesting to point out that the required number of events for the discrimination with a
properly tuned threshold-type directional detector
can be reduced to $O(10^3)$--$O(10^4)$ once the mass of the dark matters known. 
The event numbers correspond to dark matter-nucleus cross section 
$\sigma_n\simeq 10^{-30}$ cm$^2$ with exposure kg$^{-1}$day$^{-1}$ for fluorine and 
$\sigma_n\simeq 10^{-28}$ cm$^2$ for silver, respectively. 

We also analyze the case that neither the anoisotropy nor the dark matter mass are unknown.
In that case, the advantage of analysis using both recoil energy and scattering angle becomes apparent
compared to analysis using only either one of them.

\section*{Acknowledgments}
This work was supported in part by JSPS Grant-in-Aid for Scientific Research (A) 16H02189, 
for Scientific Research on Innovative Areas 26104005
and for Young Scientists (B) 26800151. KIN was also supported by Wesco research grant. KIN is grateful to M.~M.~Nojiri for fruitful discussion in the early stage of the study
and S.~Matsumoto for his keen suggestions.

\end{document}